\documentclass{ieeeaccess}
\usepackage{cite}
\usepackage{amsmath,amssymb,amsfonts}
\usepackage{algorithmic}
\usepackage{graphicx}
\usepackage{textcomp}
\usepackage{stfloats}
\usepackage[hyphens]{url}
\usepackage{verbatim}
\usepackage[table,xcdraw]{xcolor}
\usepackage{multicol}
\usepackage{multirow}
\usepackage[caption=false,font=normalsize,labelfont=sf,textfont=sf]{subfig}
\usepackage{float}

\def\BibTeX{{\rm B\kern-.05em{\sc i\kern-.025em b}\kern-.08em
    T\kern-.1667em\lower.7ex\hbox{E}\kern-.125emX}}
\begin{document}
\history{Date of publication xxxx 00, 0000, date of current version xxxx 00, 0000.}
\doi{}

\title{An Overview of UWB Standards and Organizations (IEEE 802.15.4, FiRa, Apple): Interoperability Aspects and Future Research Directions}
\author{\uppercase{Dieter Coppens}\authorrefmark{1}, \uppercase{Adnan Shahid}\authorrefmark{2},
\IEEEmembership{Senior Member, IEEE}, \uppercase{Sam Lemey}\authorrefmark{3}, \uppercase{Ben Van Herbruggen}\authorrefmark{4}, \uppercase{Chris Marshall}\authorrefmark{5}, \uppercase{Eli De Poorter}\authorrefmark{6}}
\address[1,2,3,4,6]{IDLab, Department of Information Technology, Ghent University, imec, 9052 Ghent,
Belgium}
\address[5]{imec - The Netherlands, 5656 AE Eindhoven}
\tfootnote{}

\markboth
{Dieter Coppens \headeretal: An Overview of UWB Standards and Organizations: Interoperability and Future Research}
{Dieter Coppens \headeretal: An Overview of UWB Standards and Organizations: Interoperability and Future Research}

\corresp{Corresponding author: Dieter Coppens (e-mail: dieter.coppens@ugent.be).}

\begin{abstract}
The increasing popularity of ultra-wideband (UWB) technology for location-based services, such as access control and real-time indoor track\&tracing, as well as UWB support in new consumer devices such as smartphones, has resulted in the availability of multiple new UWB radio chips. However, due to this increase in UWB device availability, the question of which (industry) standards and configuration factors impact UWB interoperability and compatibility becomes increasingly important. In this paper, the fundamentals of UWB compatibility are
investigated by first giving an overview of different UWB radio chips on the market. After that, an overview of UWB standards and organizations is given. Next, this overview is used to discuss the focus of these different standards and to identify the differences between them. We describe compatibility issues and associated interoperability aspects related to physical (PHY), medium-access-control (MAC) and upper layers. For the PHY layer, compatibility is possible for all UWB radio chips if the correct settings are configured. For the MAC layer, the implementation of the multiple access scheme as well as the localization technique is mostly proprietary. For the device discovery, several standards are currently being drafted. Finally, future challenges related to UWB interoperability are discussed.
\end{abstract}


\titlepgskip=-15pt

\maketitle

\section{Introduction}
\PARstart{U}{WB} is a general term for radio communication that uses a bandwidth close to or greater than 500 MHz \cite{ITUR}. Recently, UWB research has focused on Impulse Radio UWB (IR-UWB). This technique uses radio frequency pulses with a very short time-duration (nano- or picoseconds), resulting in a large bandwidth. The IR-UWB technique has three main benefits. The first benefit is that UWB supports a high channel capacity, due to the high bandwidth, this in turn enables the low transmission power that is needed to avoid narrowband interference with other wireless technologies. Second, the short time-duration of the pulses causes the influence of multipath to become less important, as the arrival of the pulses can be separated and filtered at the receiver. This means that UWB is robust to multipath effects, and the spatial diversity it offers can even be exploited to improve
the localization accuracy in some cases \cite{proceedingsmultipath}. The third benefit is that the high temporal resolution allows timing to be much more precise. Due to the rising edge being very steep, the receiver can very accurately determine the time of arrival of the signal, allowing centimeter-level accurate ranging using techniques such as Time-of-Flight (ToF), Time Difference of Arrival (TDoA) and Two-Way Ranging (TWR). Combining this with error correction techniques, the ranging error can be as low as 58 mm \cite{errorcorrection}. However, there are also some disadvantages. First, the low transmission power that is required to avoid narrowband interference causes UWB communication to be limited to relatively short distances. Second, the high bandwidth causes the UWB pulses to be severely distorted compared to narrowband. This can limit the performance of UWB receivers \cite{nekoogar2005ultra}.

UWB systems have received significant media attention in recent years as numerous companies, across different industries, have started adding the technology to their products. The Samsung Galaxy Note 20 Ultra contains an UWB chip that can be used in the device-to-device service called “Nearby Share” and as a digital key to unlock a door \cite{samsunguwb}. Apple iPhones use UWB to add spatial awareness to enable Apple devices to precisely locate one another \cite{appleuwb}. UWB ranging has been used for contact tracing and social distancing \cite{Loposuwb} and car manufacturers like BMW and Audi have added UWB technology for hands-free access control to their vehicles \cite{bmwuwb}.
As more UWB systems and radio chips become available, the problem of compatibility and interoperability increases. Not all UWB radio chips are open to developers, and they can support different standards, limiting the possible applications to only being available between devices using the same UWB radio chip. This can reduce the ability of this technology to reach its full potential in all applications.

To clarify the current compatibility situation, this paper explores the fundamentals of UWB compatibility. For this, different UWB standards are discussed and compared to identify the differences between them. This information is used to determine the possibilities for compatibility between two different UWB radio chips. The main contributions of the paper are the following:
\begin{enumerate}
    \item Gives a clear overview of the most prominent UWB standards.
    \item Provides a comprehensive overview of the differences between the IEEE 802.15.4 and IEEE 802.15.4z standards for both the HRP and LRP UWB PHY.
    \item Discusses the implications in hardware compatibility of the PHY, MAC and upper layers.
    \item Discusses the associated research challenges on the PHY, MAC and upper layers related to compatibility.
\end{enumerate}

The remainder of this paper is structured as follows. First, Section \ref{sec:relatedwork} reviews papers regarding UWB standards and compatibility. Second, Section \ref{sec:standards} gives an overview of the UWB standards that exist. Section \ref{sec:market} gives an overview of the different UWB radio chip market and indicates the supported standards. Next, Section \ref{sec:PHYcomp} deals with the differences between the PHY layer standards and the implications for compatibility. Section \ref{sec:MAClayer} covers the compatibility on the MAC layer. The available localization techniques and their influence is covered in Section \ref{subsec:rangingschemes}. Next, Section \ref{sec:DeviceDiscovery} describes the differences between the different standards on the subject of device discovery. In Section \ref{ResearchDirections} future research directions are given for all different layers. Finally, Section \ref{sec:conclusion} concludes the paper and discusses the lessons learned.

\section{Review of papers on UWB standards}
\label{sec:relatedwork}
In this section, recent papers on UWB and UWB standards are reviewed. First, we discuss papers focusing on the UWB PHY layer and standards, next on the MAC and upper layers. An overview of the focus of the discussed papers is also given in Table \ref{tab:related-work}.
Most UWB overview papers focus on PHY layer aspects. The authors of \cite{Sedlacek2019} give an overview of the IEEE 802.15.4z standard by looking into the changes that have been made to improve upon limitations of the IEEE 802.15.4 standard. The paper describes the improved ranging, improved timestamp robustness, improved security and reduced on-air transmission in more detail in a technical way, while also providing examples of how these features can be used. Finally, the enhancements are compared to the previous standards based on radio capabilities, ranging features and security. The main enhancements compared to the IEEE 802.15.4 standards were found to be improved first path detection, enhanced reliability of the measurement, and the new ciphered message for increasing security.

Another relevant publication is \cite{Firawhitepaper} from the FiRa consortium that first provides an overview of the development and standardization of UWB systems and technical aspects of the IEEE 802.15.4 standard. Then, the improvements made by the 802.15.4z standard are discussed similar to the previous paper but less focused on the technical aspects and the associated improvements. The second part of the paper explains the basic workings of a physical access system, the desired seamless access experience and how UWB technology can enable it. By doing this, the paper proposes methods for device discovery and other functions on higher layers than the PHY layer. 

Similar to the two previous papers about the PHY layer, the authors in \cite{singh2021security} review the most relevant concepts behind IR-UWB and the IEEE 802.15.4a and IEEE 802.15.4z standards. The difference is that the focus in this paper is on the impact on the security. The paper thus covers the enhancements made compared to the IEEE 802.15.4z standard and most importantly how they affect the security of the ranging. This is done by reviewing existing attacks and proposing new ones. This analysis shows that the IEEE 802.15.4z standard is a considerable improvement in terms of security, but securing High-Rate Pulse (HRP) ranging causes difficult trade-offs between the security and ranging performance.

The authors in \cite{Niemela2017} present an updated survey for the period of 2007 to 2015 on research related to UWB communications. In this survey, the UWB PHY layer specifications of the then two existing standards - IEEE 802.15.4-2015 and IEEE 802.15.6-2012 are discussed. A similar publication is \cite{Sharma2020} in which an overview of the standards applicable to UWB technology is given. The IEEE standards 802.15.4-2015 and 802.15.6-2012 have been compared based on modulation techniques,  interleaving,  coding  techniques  and  number  of  physical  channels. 

There are fewer overview papers that discuss UWB standardization at MAC layers. The authors of \cite{Gupta2007} study the influence of  the unique physical properties of UWB on MAC protocols for existing narrowband technologies. The media access by multiple users is addressed by reviewing MAC protocols like Carrier Sense Multiple Access with Collision Avoidance (CSMA/CA), IEEE 802.11 and IEEE 802.15.3 for UWB systems. The paper concludes that these are unsuitable for UWB systems and that further research was necessary to develop suitable MAC protocols for UWB systems. Similarly, \cite{Karapistoli2012} outlines the issues related to MAC layer design for UWB systems by highlighting the advantages and disadvantages of different MAC protocols for UWB networks.

The unique UWB physical properties do not only influence the MAC protocol design, but also provide the ability for ranging. An overview of different ranging possibilities is given in multiple scientific publications, although most of these do not address compatibility issues. For example, \cite{Alarifi2016} compares different indoor positioning technologies by comparing their performance for different metrics like, accuracy, availability, cost, coverage area and privacy. The comparison showed clearly that UWB is a promising technology for indoor positioning, mainly because of the high accuracy combined with low power usage and high level of multipath resolution. Therefore, an analysis of strengths, weaknesses, opportunities, and threats (SWOT) to analyze the present state of the UWB positioning technology is performed. An overview of different UWB positioning algorithms is given as well. Reference \cite{Shi2016} is also related to localization techniques. Here, the concept, standardization, and advantages of UWB for localization techniques are introduced. From this paper, there can be seen that UWB is a promising technology for high-accuracy indoor positioning, mainly due to its large bandwidth and low power. Four different localization techniques for UWB technology are discussed and analyzed, namely ToF, TDoA, Angle-of-Arrival (AoA) and Received Signal Strengths (RSS). This analysis found that the main advantages of AoA are that no synchronization and fewer measuring units are required in comparison to ToF and TDoA. The disadvantage is the complexity of the hardware. ToF, TDoA and AoA have a common drawback in that the performance can drop in non-line-of-sight (NLOS) situations. Using fingerprinting, RSS can perform great in such situations. The drawback is that a radio map of the indoor environment through RSS measurements needs to be created.

\begin{table*}[]
\centering
\caption{Summary of papers and survey papers on UWB and UWB standardization.}
\label{tab:related-work}
\begin{tabular}{cccccc}
\hline
 & \multicolumn{4}{c}{General aspects and standards} &  \\ \cline{2-5}
Paper & PHY layer & MAC layer & \begin{tabular}[c]{@{}c@{}}Localization \\ techniques\end{tabular} & Upper layers & \begin{tabular}[c]{@{}c@{}}Compatibility \\ of standards\end{tabular} \\ \hline
\cite{Sedlacek2019} & \checkmark&  &  &  &  \\
\cite{Firawhitepaper} & \checkmark&  &  & \checkmark&  \\
\cite{singh2021security} & \checkmark&  &  &  &  \\
\cite{Niemela2017} & \checkmark&  &  &  &  \\
\cite{Gupta2007} &  & \checkmark&  &  &  \\
\cite{Shi2016} &  &  & \checkmark&  &  \\
\cite{Karapistoli2012} &  & \checkmark&  &  &  \\
\cite{Alarifi2016} &  &  & \checkmark&  &  \\
\cite{Sharma2020} & \checkmark& \checkmark&  &  &  \\ \hline
This paper & \checkmark& \checkmark& \checkmark& \checkmark& \checkmark\\ \hline
\end{tabular}
\end{table*}

In Table \ref{tab:related-work} a summary of the focus areas of different overview papers is given. The main gap this paper tries to fill is indicated. No other scientific papers discuss the full protocol stack (PHY, MAC and upper layers) and discuss the compatibility of the different standards that are defined. Compatibility, however, is becoming increasingly important as in recent years the widespread deployment of UWB ranging systems has begun. This paper investigates the consequences for compatibility in terms of communication and ranging between UWB radio chips that do not support the same standards (or version of standards). At the same time, compatibility on MAC and upper layers is discussed as well. 

\section{Overview of UWB standards}
\label{sec:standards}
\begin{figure*}[]
    \centering
    \includegraphics[width=0.9\textwidth]{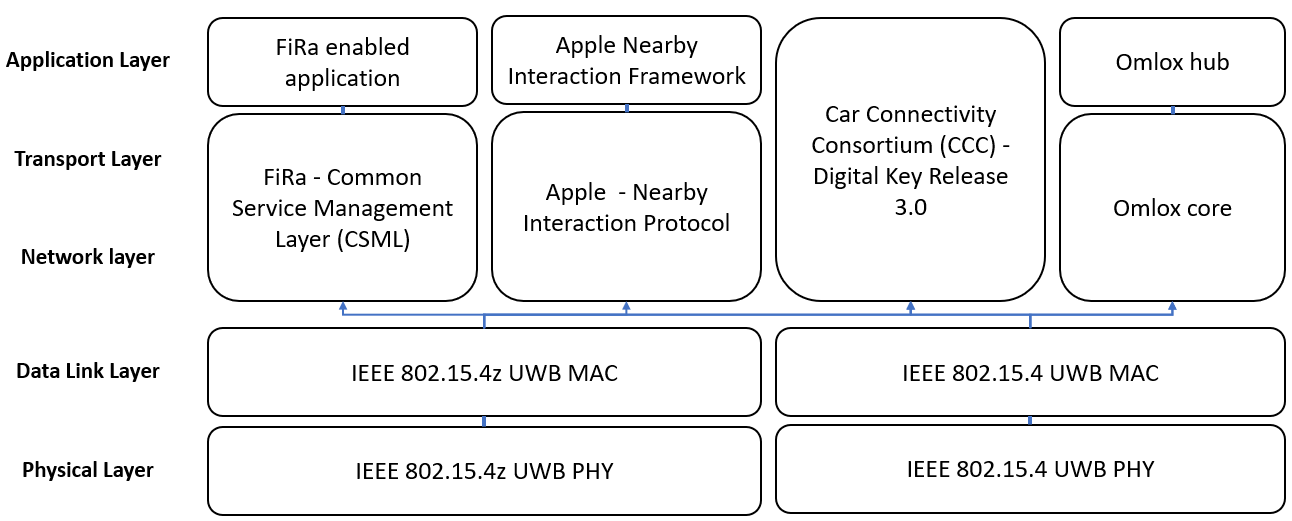}
    \caption{Overview of UWB standards and their position in the OSI reference stack.}
    \label{fig:OSI-stack}
\end{figure*}

Several organizations have defined standards for UWB. Due to the distinct roles and goals of these organizations, their standards serve different purposes. The standards are located at different layers of the Open Systems Interconnection model (OSI model). Figure \ref{fig:OSI-stack} gives an overview of the most prominent UWB standards, which will be discussed in this section, using the OSI model.
Figure \ref{fig:OSI-stack} shows that there are several standards defined for each layer of the OSI model. The presence of these different standards can complicate the compatibility, as not all UWB systems will support the same standards. This can cause UWB ranging to not being available between all UWB capable devices because compatibility between standards is not guaranteed.

\subsection{IEEE}
The IEEE Standards Association (SA) is a group within IEEE that develops global standards for a broad range of industries. IEEE SA tries to enable a neutral platform for the consensus-based development of standards by individual and corporate members \cite{IEEESA}.  
\subsubsection{IEEE 802.15.4}
The starting point for UWB standardization is the IEEE 802.15.4 standard \cite{IEEE4} that defines the MAC and PHY layers. In 2007, a first standardization of UWB technology, similar to current use of UWB technology, was provided in the IEEE 802.15.4a amendment. In this standard, UWB PHY became an IR-UWB technology focusing on low-data-rate wireless communication and especially precision ranging. In 2011, this amendment was incorporated into the main body of the standard. The IEEE 802.15.4f-2012 amendment specifies an additional UWB PHY called Low-Rate Pulse (LRP) UWB PHY. In 2015, the IEEE 802.15.4f-2011 was incorporated into the main body of the standard. This version specifies two UWB PHY modes: a) HRP and b) LRP. The HRP mode corresponds with the UWB PHY specification in IEEE 802.15.4-2011 and the LRP mode corresponds with the IEEE 802.15.4f-2012 amendment. As the name implies, the HRP mode transmits pulses at a higher rate than the LRP mode. For both LRP and HRP, the maximum transmitted energy is the same, as it is limited by the maximum mean Power Spectral Density (PSD). As a result, the HRP mode transmits more but weaker pulses and the LRP mode transmits less but stronger pulses \cite{3dbAccess,keysight}. In the remainder of the paper, we will refer to the IEEE 802.15.4-2015 version of the standard as IEEE 802.15.4.

\subsubsection{IEEE 802.15.4z}
In 2020, the IEEE 802.15.4z UWB PHY enhancement \cite{IEEE4z} to the IEEE 802.15.4 standard was released. The two main objectives of the enhancement are increasing the integrity and increasing the accuracy of ranging measurements. The enhancements include additional coding and preamble options, containing proportionally smaller sets of zero-valued elements, resulting in improved detection performance. As well as improvements to existing modulations, allowing a better balance between airtime per data bit and the number of pulses per data bit.
\begin{figure*}[]
    \centering
    \includegraphics[width=0.95\textwidth]{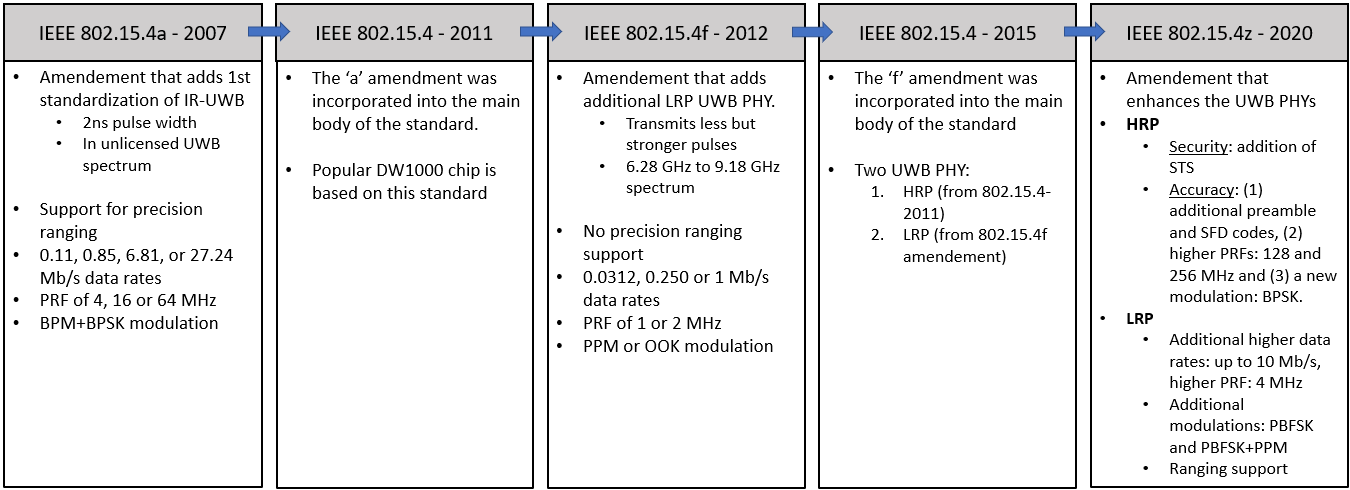}
    \caption{An overview of the changes to the IEEE 802.15.4 standard from 2007 to 2020 (based on \cite{keysight})}
    \label{fig:ieeehistory}
\end{figure*}
In Figure \ref{fig:ieeehistory} an overview of the changes made to the IEEE 802.15.4 standard related to UWB is given. The IEEE 802.15.4z merged in the standard and the new version is now called IEEE 802.15.4z-2020. In the remainder of the paper, we refer to this version as IEEE 802.15.4z.
\subsection{FiRa standard}
FiRa is an industry consortium that tries to provide a way for a wide range of product and solution companies to solve ecosystem and interoperability challenges that still occur within UWB applications.

It aims to provide a complete technical solution for UWB-services. For this, it develops profiles on top of the IEEE defined protocol layers: “FiRa is preparing a Common Service Management Layer (CSML) specification” \cite{FiRaCSML}. This is a critical specification that enables interoperability among FiRa devices and provides the framework and components needed for deploying service applications \cite{firatechspec}. 

\subsection{Apple Nearby Interaction}
Apple was one of the first companies to add UWB technology to their products. To enable third-party accessories to interact with the UWB chip in their products, Apple defined device specifications and a protocol. Companies who want to develop a UWB solution that is interoperable with the Apple UWB chip need to be part of their Made-For-iPhone (MFi) program and follow this protocol \cite{MFI}.

The Nearby Interaction Accessory Protocol Specification \cite{AppleNIprotocol} defined by Apple is a lightweight, transport-agnostic application-level protocol that enables easier configuring, starting, and maintaining of an UWB ranging session between an accessory and an Apple device.
\subsection{Car Connectivity Consortium - Digital Key 3.0}
The Car Connectivity Consortium (CCC) is a cross-industry organization with the purpose of advancing global technologies for smartphone-to-car connectivity. CCC members include car manufacturers, automotive suppliers, phone manufacturers, semiconductor suppliers, and app developers.

The Digital Key 3.0 standard \cite{CCC-digital} from the CCC implements UWB connectivity for hands-free, location-aware keyless access and location-aware features for cars. The standard ensures the highest security in localizing the device relative to the vehicle and thereby enabling the authorization of the user to access and drive the vehicle.

\subsection{Omlox}
Omlox is a collaboration within the industry in which more than 60 companies have contributed to the development. Any company can use the standard because the omlox interfaces are freely available.

Omlox is an open standard for Real-Time Location Services (RTLS). The omlox hub provides standardized interfaces for retrieving location information from a wide variety of localization techniques, such as UWB, RFID, 5G, BLE, Wi-Fi, and GPS. Information is retrieved using standardized data representations. Web-service-based instructions for finding location providers (such as mobile tags), retrieving their location and advertising new locations are defined. In addition, omlox also specifies the omlox core. The omlox core provides standardized interactions for UWB based RTLS systems, currently supporting reverse TDoA and ToF but not TDoA or TWR. The omlox core works as a possible input to the omlox hub and enables networking across UWB products, regardless of the manufacturer \cite{omlox}.

\subsection{Other UWB standards}
Currently, ranging is the most successful and widely used application of UWB technology. All previously mentioned standards and protocols are ranging related. However, there are other (either non-ranging related, electromagnetic compatibility related or based on IEEE 802.15.4) UWB standards. These standards will be discussed in this part, but will not be discussed further.
\subsubsection{IEEE 802.15.6}
The IEEE 802.15.6 is a standard for Wireless Body Area Networks (WBAN), it specifies short-range, wireless communications close to, or inside, a human body (but not limited to humans). The standard contains two UWB PHYs: IR-UWB and frequency modulated UWB (FM-UWB). The IR-UWB has similarities with the IEEE 802.15.4 mostly related to the waveform, symbol structure and frequency band allocation. The biggest difference for current applications is that in IEEE 802.15.6 the ranging protocol is not defined \cite{802156standard,Niemela2017}.

\subsubsection{IEEE 802.15.8}
This standard defines mechanisms for wireless personal area networks (WPANs) peer aware communications (PAC) The standard aims to enable scalable, low power, highly reliable wireless communications for emerging services such as social networking, advertising, gaming, streaming, and emergency services. The standard specifies two complementary UWB PHYs: (1) a Burst Position Modulation and Binary Phase-Shift Keying (BPM-BPSK) IR-UWB PHY based on IEEE 802.15.4a with new elements to improve performance  and (2) an on–off keying (OOK) UWB PHY. Both support precision ranging \cite{802158ieee}.

\subsubsection{ETSI UWB standards}
ETSI provides several standards for electromagnetic compatibility and radio spectrum matters (ERM) for UWB communication, tracking, presence detection and radar. These standards deal with technical requirements specifications such as sender and receiver compliance, spectrum access, maximum spectral density, etc. The ETSI standard is for regulatory approval of UWB devices, an implementer of the IEEE standard is responsible for referring to the applicable regulatory requirements. In the European Union, this is the ETSI standard for UWB devices \cite{ETSI}.

\subsubsection{ISO 24730 International standard}
The International Organization for Standardization (ISO) defines an air interface protocol for RTLS for use in asset management called ISO 24730-61. This standard intends to allow for compatibility and to encourage interoperability of products for the growing RTLS market. This standard defines air interface protocols, for which significant portions were excerpted from IEEE 802.15.4 (both HRP and LRP UWB PHY). Because this standard is mostly the same as the IEEE 802.15.4 it will not be discussed further \cite{ISO}.

\section{Overview of commercially available UWB radio chips}
\label{sec:market}
In Table \ref{tab:marketoverview}, an  overview of UWB radio manufacturers or designers, their UWB radio chips and the standards supported by each chip and company.

\begin{table*}[h!]
\centering
\caption{Overview of UWB radio manufacturers or designers, their UWB radio chips and the standards supported by each chip and company.}
\label{tab:marketoverview}
\begin{tabular}{|l|l|l|l|l|}
\hline
\rowcolor[HTML]{656565} 
{\color[HTML]{FFFFFF} \begin{tabular}[c]{@{}l@{}}Manufacturer \\or designer\end{tabular}} & {\color[HTML]{FFFFFF} Chip} & {\color[HTML]{FFFFFF} Supported standard} & {\color[HTML]{FFFFFF} \begin{tabular}[c]{@{}l@{}}Supported \\ UWB PHY\end{tabular}} & {\color[HTML]{FFFFFF} Standardization memberships} \\ \hline
Apple & U1 \cite{U1UWB}& IEEE 802.15.4z & HRP & CCC, FiRa Consortium, Apple Nearby Interaction \\ \hline
Qorvo & DW1000 \cite{DW1000uwb} & IEEE 802.15.4-2011 & / &  \\ \cline{1-4}
Qorvo & DW3000 family \cite{DW3000family}  & IEEE 802.15.4z & HRP & \multirow{-2}{*}{CCC, FiRa Consortim, Apple Nearby Interaction} \\ \hline
NXP & SR040/SR150 \footnotemark \cite{NXPSR040uwb} \cite{NXPSR150uwb} & IEEE 802.15.4z & HRP &  \\ \cline{1-4}
NXP & NCJ29D5 \footnotemark[\value{footnote}] \cite{NXPNCJuwb} & IEEE 802.15.4 & HRP & \multirow{-2}{*}{CCC, FiRa Consortium, Apple Nearby Interaction} \\ \hline
Imec & ULP IR-UWB chip \cite{imecuwb} & IEEE 802.15.4z & HRP & CCC, FiRa Consortium \\ \hline
3db Access & 3DB6830 \cite{3dbAccess} & IEEE 802.15.4 & LRP & FiRa consortium \\ \hline
Zebra technologies & Zebra UWB chip \cite{zebrauwb} & IEEE 802.15.4 & LRP &  \\ \hline
\end{tabular}
\end{table*}

\footnotetext{All information about this chip specified in the remainder of this paper has been found in public information (source specified in this case) or is derived from the supported PHY layer standards.}

When comparing the PHY standards supported by the different UWB radio chips, it can be seen that the most widely supported standard is the IEEE 802.15.4z standard. The DW1000, however, only supports the IEEE 802.15.4 HRP. This is important for compatibility, as the DW1000 is the most widely used chip for research purposes and is also used in numerous commercial products that provide RTLS.
The UWB chip from 3dB access and Zebra technologies are the only two UWB radio chips supporting the LRP mode of a standard. In section \ref{sec:PHYcomp} the difference with this standard will be discussed in more detail. 

\section{PHY compatibility}
\label{sec:PHYcomp}
This section discusses the PHY configuration and compatibility challenges in more detail. There are two main PHY standards defined: the IEEE 802.15.4 and IEEE 802.15.4z standard. As mentioned in Section \ref{sec:standards}, the IEEE 802.15.4z standard is an enhancement to the IEEE 802.15.4 standard, with the two main objectives being increasing the security and increasing the accuracy of ranging measurements. The increased security is necessary to enable safe hands-free access control applications, as security is extremely important to stop attackers from getting access to buildings or cars using UWB technology. The accuracy is increased to improve the ranging performance. When supporting the IEEE standard, it is not mandatory to support every feature from that standard. This means that there can be differences between UWB radio chips supporting the same standard. For UWB radio chips to be compatible on the PHY layer, there are several conditions that must be met: the pulse shape needs to be similar, the used center frequency and frame structure needs to be same. This section will address these conditions based on the different UWB radio chips discussed in Section \ref{sec:market}.

The IEEE UWB PHY consists of two modes: HRP and LRP. A comparison between the HRP and LRP UWB PHY features is shown in Table \ref{tab:LRPvsHRP}. HRP was the only mode used for ranging in the IEEE 802.15.4 standard. Because of this, it is the most widely used mode in commercial UWB products and chips. The IEEE 802.15.4z enhancement enables ranging in the LRP UWB by implementing a basic ranging scheme.

The UWB chip from 3dB access and Zebra technologies are based on the LRP UWB, all other UWB radio chips in Table \ref{tab:marketoverview} only support the HRP UWB PHY. As a result, compatibility between the other UWB radio chips is not possible. In the next subsections, compatibility within the individual PHYs is discussed \cite{IEEE4,IEEE4z}.

\begin{table*}[]
\centering
\caption{Overview of HRP UWB and LRP UWB features and the changes in the IEEE 802.15.4z \cite{IEEE4,IEEE4z}. }
\label{tab:LRPvsHRP}
\begin{tabular}{|c|cc|cc|}
\hline
\rowcolor[HTML]{656565} 
{\color[HTML]{FFFFFF} } & \multicolumn{2}{c|}{\cellcolor[HTML]{656565}{\color[HTML]{FFFFFF} LRP UWB}} & \multicolumn{2}{c|}{\cellcolor[HTML]{656565}{\color[HTML]{FFFFFF} HRP UWB}} \\ \hline
\rowcolor[HTML]{656565} 
{\color[HTML]{FFFFFF} Feature} & \multicolumn{1}{c|}{\cellcolor[HTML]{656565}{\color[HTML]{FFFFFF} \begin{tabular}[c]{@{}c@{}}IEEE \\ 802.15.4\end{tabular}}} & {\color[HTML]{FFFFFF} \begin{tabular}[c]{@{}c@{}}Added in\\IEEE 802.15.4z\end{tabular}} & \multicolumn{1}{c|}{\cellcolor[HTML]{656565}{\color[HTML]{FFFFFF} \begin{tabular}[c]{@{}c@{}}IEEE\\ 802.15.4\end{tabular}}} & {\color[HTML]{FFFFFF} \begin{tabular}[c]{@{}c@{}}Added in\\IEEE 802.15.4z\end{tabular}} \\ \hline
Data rates & \multicolumn{1}{c|}{\begin{tabular}[c]{@{}c@{}}31.25 kbps\\ 250 kbps\\ 1 Mbps\end{tabular}} & \begin{tabular}[c]{@{}c@{}}2,3,4,5,6 Mbps\\ 8 Mbps\\ 10 Mbps\end{tabular} & \multicolumn{1}{c|}{\begin{tabular}[c]{@{}c@{}}110 kbps\\ 850 kbps\\ 6.81 Mbps\\ 27.24 Mbps\end{tabular}} &  \\ \hline
\rowcolor[HTML]{C0C0C0} 
Peak pulse repetition rate & \multicolumn{1}{c|}{\cellcolor[HTML]{C0C0C0}2 MHz} & 4 MHz & \multicolumn{1}{c|}{\cellcolor[HTML]{C0C0C0}499.2 MHz} &  \\ \hline
Ranging support & \multicolumn{1}{c|}{No} & Yes & \multicolumn{1}{c|}{Yes} &  \\ \hline
\rowcolor[HTML]{C0C0C0} 
\begin{tabular}[c]{@{}c@{}}Multi-user interference \\ suppression\end{tabular} & \multicolumn{1}{c|}{\cellcolor[HTML]{C0C0C0}No} &  & \multicolumn{1}{c|}{\cellcolor[HTML]{C0C0C0}Yes} &  \\ \hline
Modulation & \multicolumn{1}{c|}{\begin{tabular}[c]{@{}c@{}}OOK\\ PPM\end{tabular}} & \begin{tabular}[c]{@{}c@{}}PBFSK\\ BPFSK + 8/16/32PPM\end{tabular} & \multicolumn{1}{c|}{BPSK + BPM} & BPSK \\ \hline
\rowcolor[HTML]{C0C0C0} 
Error correction & \multicolumn{1}{c|}{\cellcolor[HTML]{C0C0C0}\begin{tabular}[c]{@{}c@{}}SECDED\\ Convolutional\end{tabular}} &  & \multicolumn{1}{c|}{\cellcolor[HTML]{C0C0C0}\begin{tabular}[c]{@{}c@{}}SECDED\\ Convolutional (K=3)\\ Reed-Solomon\end{tabular}} & Convolutional (K=7) \\ \hline
\end{tabular}
\end{table*}

\subsection{HRP UWB PHY compatibility}
\subsubsection{Channel}
The first condition for two UWB radio chips to be compatible is that they need to be able to use the same center frequency and bandwidth. Without this, no reception is possible. The IEEE 802.15.4/4z HRP standards define the same 16 channels or bands, each channel is a combination of a center frequency and a maximum bandwidth. The allocation is shown in Table \ref{tab:bandallocation}. It can be seen that the minimum bandwidth is 499.2 MHz and that some channels have the same center frequency but a different bandwidth. This is the case for channels 2 and 4, channels 5 and 7, channels 9 and 11 and lastly for channels 13 and 15.

\begin{table}[h!]
\centering
\caption{HRP UWB  band allocation \cite{IEEE4}.}
\label{tab:bandallocation}
\begin{tabular}{|c|c|c|}
\hline
\rowcolor[HTML]{656565} 
{\color[HTML]{FFFFFF} \textbf{Channel number}} & {\color[HTML]{FFFFFF} \textbf{Center frequency (MHz) }} & {\color[HTML]{FFFFFF} Bandwidth (MHz)} \\ \hline
{\color[HTML]{333333} 0} & {\color[HTML]{333333} 499.2} & 499.2 \\ \hline
\rowcolor[HTML]{C0C0C0} 
{\color[HTML]{333333} 1} & {\color[HTML]{333333} 3494.4} & 499.2 \\ \hline
\rowcolor[HTML]{FFFFFF} 
{\color[HTML]{333333} 2} & {\color[HTML]{333333} 3993.6} & 499.2 \\ \hline
\rowcolor[HTML]{C0C0C0} 
{\color[HTML]{333333} 3} & {\color[HTML]{333333} 4992.8} & 499.2 \\ \hline
\rowcolor[HTML]{FFFFFF} 
{\color[HTML]{333333} 4} & {\color[HTML]{333333} 3993.6} & 1331.2 \\ \hline
\rowcolor[HTML]{C0C0C0} 
{\color[HTML]{333333} 5} & {\color[HTML]{333333} 6489.6} & 499.2 \\ \hline
6 & 6988.8 & \cellcolor[HTML]{FFFFFF}499.2 \\ \hline
\rowcolor[HTML]{C0C0C0} 
7 & 6489.6 & 1081.6 \\ \hline
8 & 7448.0 & \cellcolor[HTML]{FFFFFF}499.2 \\ \hline
\rowcolor[HTML]{C0C0C0} 
9 & 7987.2 & 499.2 \\ \hline
10 & 8486.4 & \cellcolor[HTML]{FFFFFF}499.2 \\ \hline
\rowcolor[HTML]{C0C0C0} 
11 & 7987.2 & 1331.2 \\ \hline
12 & 8985.6 & \cellcolor[HTML]{FFFFFF}499.2 \\ \hline
\rowcolor[HTML]{C0C0C0} 
13 & 9494.8 & 499.2 \\ \hline
14 & 9984.0 & \cellcolor[HTML]{FFFFFF}499.2 \\ \hline
\rowcolor[HTML]{C0C0C0} 
15 & 9484.8 & 1354.97 \\ \hline
\end{tabular}
\end{table}

While the HRP PHY contains 16 different channels, (see Table \ref{tab:bandallocation}) it is not mandatory to support every channel. For the sub-gigahertz operation, channel 0 is the only mandatory channel; for the low-band operation, channel 3 is the mandatory channel; and for the high-band operation, channel 9 is the mandatory channel. This means that not all UWB radio chips support the same channels. An overview of the channels that are supported by each chip is given in Table \ref{tab:channel}. For the NXP NCJ29D5 chip, the supported channels are not explicitly published, but a 6-8 GHz band operation is specified. the non-mandatory channels in this band are indicated with a question mark. For the imec ULP IR-UWB chip, the possible channels are indicated. However, due to the chip being a design, there could be differences in actual implementations of this chip. Imec sells design information which manufacturers use in their UWB radio chips. This means that final decisions in which features are supported are not made by imec, but by the manufacturer using the design information. Whether a feature is supported or not can be decided by the product management of the manufacturer for different reasons, like chip area, current consumption, time to market, specification stability, test requirements, software support, etc. This is also the case for other features discussed below. This table clearly indicates that all UWB radio chips mentioned in the market overview support channel 5. This indicates that this aspect of compatibility can always be fulfilled by using channel 5. All UWB radio chips, except for the Qorvo DW1000, support channel 9 as well.

\begin{table*}[h!]
\caption{Overview of the channels supported by each chip.}
\label{tab:channel}
\centering
\begin{tabular}{|l|l|l|l|l|l|l|l|l|l|l|l|}
\hline
\rowcolor[HTML]{656565} 
\multicolumn{1}{|c|}{\cellcolor[HTML]{656565}{\color[HTML]{FFFFFF} \begin{tabular}[c]{@{}l@{}}Manufacturer \\ or designer\end{tabular}}} & \multicolumn{1}{c|}{\cellcolor[HTML]{656565}{\color[HTML]{FFFFFF}\textbf{Chip} }} & \multicolumn{10}{c|}{\cellcolor[HTML]{656565}{\color[HTML]{FFFFFF} \textbf{Channel index}}} \\ \cline{3-12} 
\rowcolor[HTML]{656565} 
\multicolumn{1}{|c|}{\multirow{2}{*}{\cellcolor[HTML]{656565}{\color[HTML]{FFFFFF} }}} & \multicolumn{1}{c|}{\multirow{2}{*}{\cellcolor[HTML]{656565}{\color[HTML]{FFFFFF} }}} & {\color[HTML]{FFFFFF} 0} & {\color[HTML]{FFFFFF} 1} & {\color[HTML]{FFFFFF} 2} & {\color[HTML]{FFFFFF} 3} & {\color[HTML]{FFFFFF} 4} & {\color[HTML]{FFFFFF} 5} & {\color[HTML]{FFFFFF} 6} & {\color[HTML]{FFFFFF} 7} & {\color[HTML]{FFFFFF} 8} & {\color[HTML]{FFFFFF} 9} \\ \hline
{\color[HTML]{333333} Apple} & {\color[HTML]{333333} U1 \cite{U1techinsights}} &  &  &  &  &  & \checkmark&  &  &  & \checkmark\\ \hline
\rowcolor[HTML]{C0C0C0} 
{\color[HTML]{333333} Qorvo} & {\color[HTML]{333333} DW1000} &  & \checkmark& \checkmark& \checkmark& \checkmark& \checkmark&  &  &  &  \\ \hline
{\color[HTML]{333333} Qorvo} & {\color[HTML]{333333} DW3000 Family} &  &  &  &  &  & \checkmark&  &  &  & \checkmark\\ \hline
\rowcolor[HTML]{C0C0C0} 
{\color[HTML]{333333} NXP} & {\color[HTML]{333333} SR040/SR150 \cite{SR040} \cite{SR150}} &  &  &  &  &  & \checkmark& &  &  & \checkmark\\ \hline
{\color[HTML]{333333} NXP} & {\color[HTML]{333333} NCJ29D5} &  &  &  &  &  & \checkmark& ? & ? & ?& \checkmark\\ \hline
\rowcolor[HTML]{C0C0C0} 
{\color[HTML]{333333} Imec} & {\color[HTML]{333333} ULP IR-UWB   radio} &  &  &  &  &  & \checkmark& \checkmark&  & \checkmark& \checkmark\\ \hline
\end{tabular}

\end{table*}

\subsubsection{Pulse shape}
\begin{figure}[ht]
    \centering
    \includegraphics[width=0.49\textwidth]{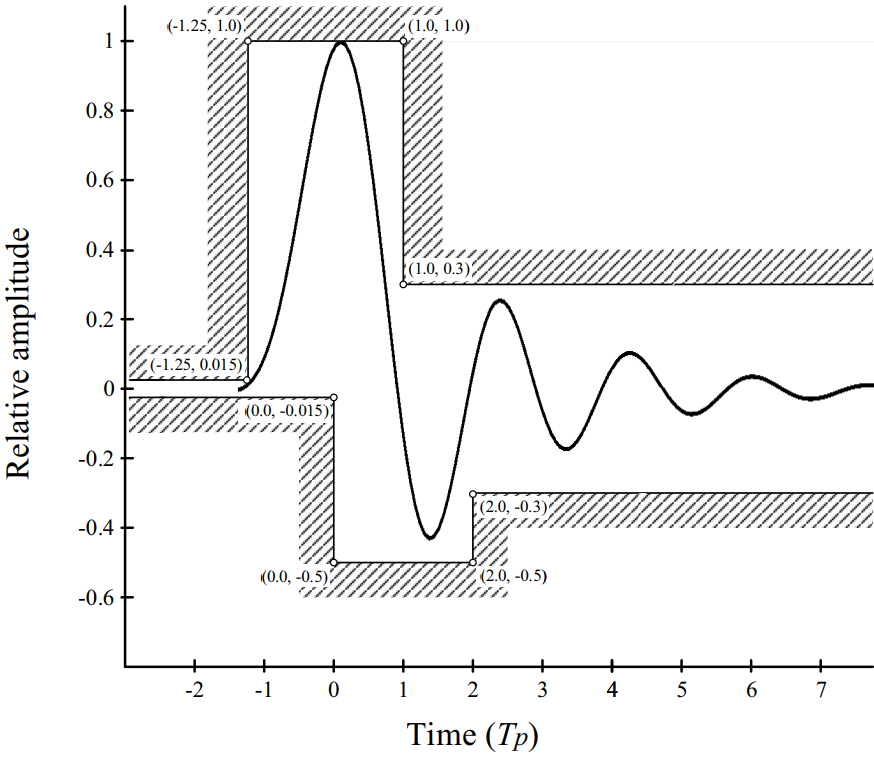}
    \caption{Time domain mask for an IEEE 802.15.4/4z HRP compliant pulse \cite{IEEE4}.}
    \label{fig:timedomainmask}
\end{figure}

The IEEE 802.15.4 standard and IEEE 802.15.4z enhancement both have the same requirements for the pulse shape in HRP UWB. The transmitted pulse shape $p(t)$ shall be constrained by the shape of its cross-correlation function with
a standard reference pulse $r(t)$. This reference pulse is a root-raised-cosine pulse with a roll-off factor of $\beta = 0.5$.
In order for a  transmitter to be compliant with the standard, the transmitted pulse p(t) needs to have a magnitude of the cross-correlation function $|\phi(\tau) |$ whose main lobe is greater than or equal to 0.8 for a duration of at least $T_w$, as defined in Table \ref{tab:referencepulse}, and all side lobes need to be smaller than 0.3.
A second requirement for the pulse is the time domain mask, shown in Figure \ref{fig:timedomainmask}. A pulse that is compliant with the standard cannot exceed the bounds that are set, this is to comply with the spectrum constraints inherited from the FCC and other regulatory bodies \cite{fcc,ETSI}.

\begin{table}[ht]
\centering
\caption{Required reference pulse durations in each channel for HRP UWB.}
\label{tab:referencepulse}
\begin{tabular}{|c|c|c|}
\hline
\rowcolor[HTML]{656565} 
{\color[HTML]{EFEFEF} \begin{tabular}[c]{@{}c@{}}Channel\\  number\end{tabular}} & {\color[HTML]{EFEFEF} \begin{tabular}[c]{@{}c@{}}Pulse duration\\ $T_p$ (ns)\end{tabular}} & {\color[HTML]{EFEFEF} \begin{tabular}[c]{@{}c@{}}Main lobe\\ width $T_w$ (ns)\end{tabular}} \\ \hline
\{0:3, 5:6,8:10,12:14\} & 2.00 & 0.50 \\ \hline
\rowcolor[HTML]{9B9B9B} 
7 & 0.92 & 0.20 \\ \hline
\{4,11\} & 0.75 & 0.20 \\ \hline
\rowcolor[HTML]{9B9B9B} 
15 & 0.74 & 0.20 \\ \hline
\end{tabular}
\end{table}

\begin{figure}[h!]
    \centering
    \includegraphics[width=0.49\textwidth]{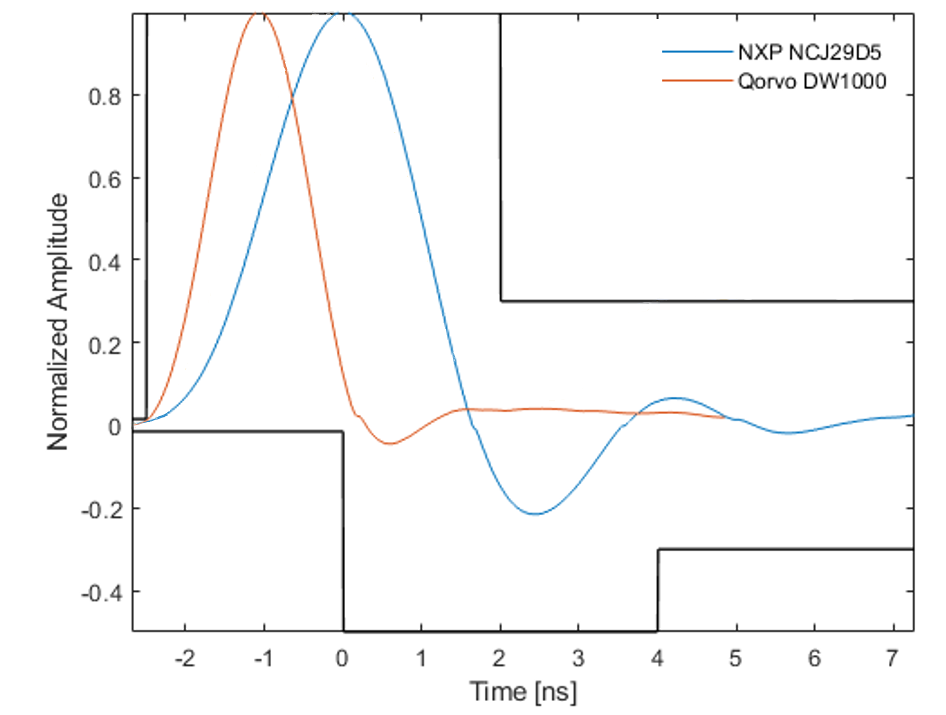}
    \caption{Normalized time-domain pulses of the Qorvo DW1000 and NXP NCJ29D5 UWB radio chips, measured using the LabMaster 10 ZI-A Oscilloscope from Teledyne Lecroy, plot together with the time-domain mask.}
    \label{fig:pulses}
\end{figure}

This should mean that the pulses, transmitted by all UWB radio chips that are compliant to the standards, are compatible. However, being compliant to the pulse requirements of the standards does not mean that the pulses are the same. Differences in pulse shape between different UWB radio chips are possible while still both being compliant. This can be seen in Figure \ref{fig:pulses}, here  the default pulse of two different UWB radio chips, namely the Qorvo DW1000 and NXP NCJ29D5, are shown. The UWB radio chips were connected to the LabMaster 10 ZI-A Oscilloscope from Teledyne Lecroy using a cable. The measurements are done in the time-domain using a sampling rate of 160 GS/s. It can be seen that it is possible for the width of the pulse to differ among compliant pulses.

\textit{Consequences of difference in pulse width}: A difference in pulse width can influence the ranging accuracy because the timing on the pulses can differ. There are multiple ways to time on a pulse, such as half-amplitude timing and or threshold crossing. If half-amplitude timing was used on the pulses shown in Figure \ref{fig:pulses}, the difference in pulse width could lead to a difference in timing of more than 0.5 ns resulting in a difference in ranging distances of more than 15 cm. This is significant as centimeter-level accuracy is expected of UWB systems. However, calibrating the antenna delay parameters in the UWB ranging systems will solve this problem.

\subsubsection{Frame structure}
\begin{figure}[h!]
    \centering
    \includegraphics[width=0.49\textwidth]{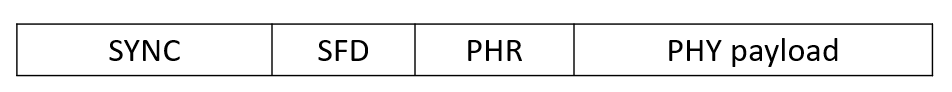}
    \caption{Structure of a HRP UWB  frame in the IEEE 802.15.4 standard.}
    \label{fig:structure-4}
\end{figure}

The HRP UWB  frame structure of the IEEE 802.15.4 standard consists of up to four different fields and is shown in Figure \ref{fig:structure-4}. 

\begin{figure}[h!]
    \centering
    \includegraphics[width=0.49\textwidth]{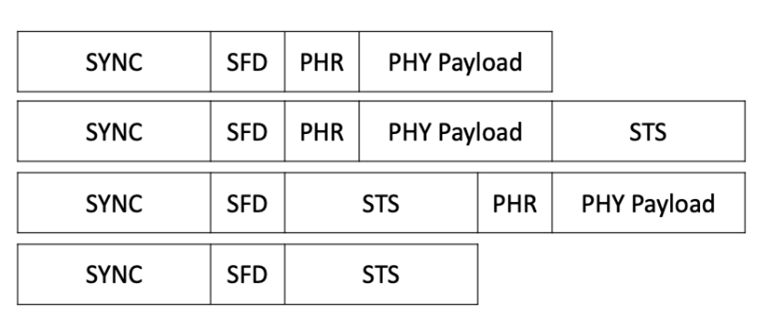}
    \caption{All four possible HRP UWB  frame structures in the IEEE 802.15.4z standard (based on \cite{IEEE4z}).}
    \label{fig:structure-4z}
\end{figure}

In IEEE 802.15.4z there are four possible frames, which consist of up to five different fields, defined as shown in Figure \ref{fig:structure-4z}. One of the four frame structures is equivalent to the frame structure of the IEEE 802.15.4 standard. The other three are different due to the addition of a new field called the Scrambled Timestamp Sequence (STS) field.

For communication to be possible between two different UWB radios, the configured frame structure needs to be the same. UWB radios that only support the IEEE 802.15.4 standard cannot use the STS field and can thus also not use the three new frame structures defined in the IEEE 802.15.4z that use this field. 
\paragraph{Synchronization (SYNC) field}
The purpose of the SYNC field or preamble is to synchronize the sender and receiver. The receiver detects the preamble and synchronizes to the sender in line with the preamble. The preamble sequence itself is constructed from a ternary code (alphabet {1,0,-1}) where 1 stands for a positive pulse, -1 for a negative pulse and 0 for no pulse. Each channel has a minimum of two compatible codes. The codes for one channel are chosen to have a low cross-correlation factor with each other. This allows multiple devices to operate using the same channel simultaneously without interference. The code is then spread to construct a symbol Si by inserting zeros between each ternary element of the code. To form the complete preamble, this symbol is repeated a number of times. This parameter is called Preamble Symbol Repetitions (PSR). In the IEEE 802.15.4 standard, there are two different ternary code lengths defined, namely 31 and 127. The IEEE 802.15.4z standard support the ternary codes with a length of 127 from the IEEE 802.15.4 standard and defines new dense (contains fewer zeros) ternary codes with a length of 91. These new ternary codes are defined to enable more accurate timing. A receiver needs to offer a high dynamic range to be able to successfully detect the direct path. In the HRP UWB, high dynamic range is obtained by correlation. As shown in \cite{UWBPARAM4z} improvement of the dynamic range is possible by increasing the number of threshold decision events. This can be done by defining preamble codes that contain fewer zeros (position where no pulse is sent). This causes more pulses to be sent and thus a higher accuracy.

\begin{table*}[h!]
\centering
\caption{Overview of supported preamble codes for each chip.}
\label{tab:preamble}
\begin{tabular}{|l|l|c|c|c|}
\hline
\rowcolor[HTML]{656565} 
\multicolumn{1}{|c|}{\cellcolor[HTML]{656565}{\color[HTML]{FFFFFF} \begin{tabular}[c]{@{}l@{}}Manufacturer \\ or designer\end{tabular}}} & \multicolumn{1}{c|}{\cellcolor[HTML]{656565}{\color[HTML]{FFFFFF} \textbf{Chip}}} & {\color[HTML]{FFFFFF} \begin{tabular}[c]{@{}l@{}}Ternary code\\ length 31\end{tabular}} & {\color[HTML]{FFFFFF} \begin{tabular}[c]{@{}l@{}}Ternary code\\ length 127\end{tabular}} & {\color[HTML]{FFFFFF} \begin{tabular}[c]{@{}l@{}}Ternary code\\ length 91\end{tabular}} \\ \hline
{\color[HTML]{333333} Apple} & {\color[HTML]{333333} U1} & ? & \checkmark& ?\\ \hline
\rowcolor[HTML]{C0C0C0} 
{\color[HTML]{333333} Qorvo} & {\color[HTML]{333333} DW1000} & \checkmark& \checkmark&  \\ \hline
{\color[HTML]{333333} Qorvo} & {\color[HTML]{333333} DW3000 Family} & \checkmark& \checkmark&  \\ \hline
\rowcolor[HTML]{C0C0C0} 
{\color[HTML]{333333} NXP} & {\color[HTML]{333333} SR040/SR150} & ? & \checkmark& ?\\ \hline
{\color[HTML]{333333} NXP} & {\color[HTML]{333333} NCJ29D5} & \checkmark& \checkmark& ?\\ \hline
\rowcolor[HTML]{C0C0C0} 
{\color[HTML]{333333} Imec} & {\color[HTML]{333333} ULP IR-UWB   radio} &  \checkmark& \checkmark& \checkmark\\ \hline
\end{tabular}
\end{table*}

In Table \ref{tab:preamble}, an overview of the preamble codes supported by each chip is given. All UWB radio chips support the ternary codes with length 127 as these codes are mandatory in both standards. The new dense ternary code with a length of 91 can be supported by all UWB radio chips supporting the IEEE 802.15.4z. The dense ternary code of length 91 is not mandatory, and the designers of the DW3000 chose not to support it. No further details are available for the Apple and NXP UWB radio chips, therefore question marks are placed if support of the preamble is possible but not certain. 

If a different code is configured at the receiver than at the sender, no communication is possible due to the UWB radio chips not being synchronized. As mentioned before, this is done by design to allow multiple devices to operate using the same channel without interference. To connect an UWB chip supporting the IEEE 802.15.4 and an UWB chip supporting the IEEE 802.15.4z, the ternary code with a length of 127 needs to be used. This means that the higher accuracy enabled by the new dense ternary codes is not available for this connection.

\paragraph{Start-of-Frame Delimiter (SFD) field}
The SFD indicates the end of the preamble and the precise start of the switch to the PHY header (PHR). The SFD is also used for timestamping and thus important for ranging performance. The IEEE 802.15.4 standard defines two ternary codes: a short SFD code with a length of 8 and a long SFD code with a length of 64. These codes are then spread by the preamble symbol $S_{i}$. A 1 indicates that the preamble symbol is repeated, -1 corresponds with the preamble being transmitted with opposite polarity  and 0 indicates that no pulses are being transmitted for the length of the preamble symbol $S_{i}$. The IEEE 802.15.4z standard drops support for the ternary code with a length of 64 and defines 4 new binary codes with a length of 4, 8, 16 and 32. The purpose of these new codes is similar to the new preamble codes. Being binary, there is no position where pulses are not sent, this leads to more pulses being transmitted and thus a higher accuracy.

\begin{table*}[h!]
\caption{Overview of supported SFD codes for each chip.}
\label{tab:SFD}
\centering
\begin{tabular}{|l|l|c|c|c|}
\hline
\rowcolor[HTML]{656565} 
\multicolumn{1}{|c|}{\cellcolor[HTML]{656565}{\color[HTML]{FFFFFF} \begin{tabular}[c]{@{}l@{}}Manufacturer \\ or designer\end{tabular}}} & \multicolumn{1}{c|}{\cellcolor[HTML]{656565}{\color[HTML]{FFFFFF} \textbf{Chip}}} & {\color[HTML]{FFFFFF} \begin{tabular}[c]{@{}l@{}}Ternary code\\ length 8\end{tabular}} & {\color[HTML]{FFFFFF} \begin{tabular}[c]{@{}l@{}}Ternary code\\ length 64\end{tabular}} & {\color[HTML]{FFFFFF} Binary codes} \\ \hline
{\color[HTML]{333333} Apple} & {\color[HTML]{333333} U1} & \checkmark&?  & \checkmark\\ \hline
\rowcolor[HTML]{C0C0C0} 
{\color[HTML]{333333} Qorvo} & {\color[HTML]{333333} DW1000} & \checkmark& \checkmark&  \\ \hline
{\color[HTML]{333333} Qorvo} & {\color[HTML]{333333} DW3000 Family} & \checkmark& \checkmark& \checkmark\\ \hline
\rowcolor[HTML]{C0C0C0} 
{\color[HTML]{333333} NXP} & {\color[HTML]{333333} SR040/SR150} & \checkmark& ?& \checkmark\\ \hline
{\color[HTML]{333333} NXP} & {\color[HTML]{333333} NCJ29D5} & \checkmark& ?& \checkmark\\ \hline
\rowcolor[HTML]{C0C0C0} 
{\color[HTML]{333333} Imec} & {\color[HTML]{333333} ULP IR-UWB   radio} & \checkmark&  & \checkmark\\ \hline
\end{tabular}
\end{table*}

\begin{figure}[h!]
    \centering
    \includegraphics[width=0.49\textwidth]{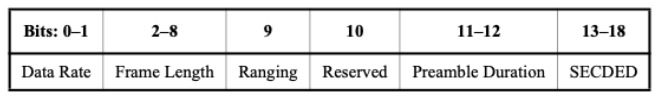}
    \caption{PHR field format in the IEEE 802.15.4 standard \cite{IEEE4}.}
    \label{fig:PHR}
\end{figure}

For communication to be possible between two UWB radio chips, the configured SFD needs to be the same, as otherwise the sender and receiver cannot determine the time of arrival correctly. In Table \ref{tab:SFD} an overview of the supported SFD codes supported by each chip is given. There can be seen that all UWB radio chips support the short ternary code, which indicates that compatibility for this aspect is possible if that short ternary SFD code is used. To connect an UWB chip supporting the IEEE 802.15.4 and an UWB chip supporting the IEEE 802.15.4z, the only SFD code that can be used is the short ternary code. This means that the higher accuracy enabled by the new binary codes is not available for this connection. In Table \ref{tab:SFD} the binary codes are grouped together in one category. The possible lengths of these codes are 4,8,16 and 32. Not all UWB radio chips support every possible length. 
\paragraph{PHR field}
After the preamble and SFD parts of the frame, the actual data held in the package will begin. This starts with a field called PHR. The purpose of this field is to give information about the payload that is transmitted to the receiver. Figure \ref{fig:PHR} shows the format of the PHR field in the IEEE 802.15.4 standard.

The following fields are present in the PHR:
\begin{itemize}
    \item \textbf{The Data Rate field}: indicates the data rate of the received PHY Payload field. The PHR is sent at 850 kbps for all data rates greater or equal than 850 kbps and at 110 kbps for the data rate of 110 kbps.
    \item \textbf{The Frame Length field}: is an unsigned integer number that indicates the number of octets in the payload.
    \item \textbf{The Ranging field}: shall be set to one if the current frame is used for ranging and shall be set to zero otherwise.
    \item \textbf{Preamble Duration field}: represents the length in preamble symbols of the SYNC field.
    \item \textbf{The Single Error Correct Double Error Detect (SECDED) field}: a simple Hamming block code that enables the correction of a single error and the detection of two errors at the receiver.
\end{itemize}

The IEEE 802.15.4z standard allows for more data to be transmitted in one packet. The PHR field can be configured to allow for the PHY payload length field to increase up to 12 bits by eliminating the preamble duration and reserved field and optionally the data rate field. This allows the maximum payload length to increase from 128 to 4096 bytes. 

From Table \ref{tab:PHR} it can be seen that all UWB radio chips support the IEEE 802.15.4 PHR format, this is due to that format being mandatory. The format to allow for increased payload length is optional. To connect an UWB chip supporting the IEEE 802.15.4 and an UWB chip supporting the IEEE 802.15.4z, only the IEEE 802.15.4 PHR can be used. This has the consequence that the maximum payload length remains at 128 bytes for such a connection.

\begin{table}[h!]
\caption{Overview of the supported PHR format for each chip.}
\label{tab:PHR}
\centering
\begin{tabular}{|l|l|c|c|}
\hline
\rowcolor[HTML]{656565} 
\multicolumn{1}{|c|}{\cellcolor[HTML]{656565}{\color[HTML]{FFFFFF} \begin{tabular}[c]{@{}l@{}}Manufacturer \\ or designer\end{tabular}}} & \multicolumn{1}{c|}{\cellcolor[HTML]{656565}{\color[HTML]{FFFFFF} \textbf{Chip}}} & \multicolumn{1}{l|}{\cellcolor[HTML]{656565}{\color[HTML]{FFFFFF} \begin{tabular}[c]{@{}c@{}}IEEE \\ 802.15.4 \\ PHR \end{tabular}}} & \multicolumn{1}{l|}{\cellcolor[HTML]{656565}{\color[HTML]{FFFFFF} \begin{tabular}[c]{@{}c@{}}IEEE \\ 802.15.4z \\ PHR \end{tabular}}} \\ \hline
{\color[HTML]{333333} Apple} & {\color[HTML]{333333} U1} & \checkmark& ?\\ \hline
\rowcolor[HTML]{C0C0C0} 
{\color[HTML]{333333} Qorvo} & {\color[HTML]{333333} DW1000} & \checkmark&  \\ \hline
{\color[HTML]{333333} Qorvo} & {\color[HTML]{333333} DW3000 Family} & \checkmark&  \\ \hline
\rowcolor[HTML]{C0C0C0} 
{\color[HTML]{333333} NXP} & {\color[HTML]{333333} SR040/SR150} & \checkmark& ?\\ \hline
{\color[HTML]{333333} NXP} & {\color[HTML]{333333} NCJ29D5} & \checkmark& ?\\ \hline
\rowcolor[HTML]{C0C0C0} 
{\color[HTML]{333333} Imec} & {\color[HTML]{333333} ULP IR-UWB radio} & \checkmark& \checkmark\\ \hline
\end{tabular}
\end{table}

\paragraph{STS field}
The amount of possible preamble codes is limited, and they are repeated several times in the SYNC field. This opens the door for attackers \cite{Poturalski2011}. To combat this, a new optional field, the STS, is inserted into the UWB frame. The presence and position of this field determines four different configurations, as shown in Figure \ref{fig:structure-4z}. The STS works like the preamble, but it does not repeat itself. It is a sequence of pseudo-randomized
pulses generated by a Deterministic Random Bit Generator (DRBG) arranged in (1 to 4) blocks of active
segments encapsulated by silent intervals or gaps. Due to the pseudo-randomness of the sequence, there is no periodicity, allowing reliable, highly accurate, and artifact-free channel estimates to be produced by the receiver. To generate the STS, the DRBG produces 128-bit pseudo-random numbers using a seed consisting of a 128-bit key, and a 128-bit nonce (a number that should only be used once). The nonce is updated during the STS generation by incrementing the counter once for every 128-bit number generated. Each bit of value zero produces a positive polarity pulse, and each bit of value one produces a negative polarity pulse. These pulses are then spread. To decode the STS, the receiver needs to have a copy of the sequence locally available before the start of reception. This is only possible if both transmitter and receiver know the keys and cryptographic scheme for STS generation. The STS cannot replace the preamble field and is always behind the SFD, since the STS correlation only works if it is started at the same time.

\begin{table}[h!]
\caption{Overview of which UWB radio chips support the use of the STS field.}
\label{tab:STSsupport}
\centering
\begin{tabular}{|l|l|l|}
\hline
\rowcolor[HTML]{656565} 
\multicolumn{1}{|l|}{\cellcolor[HTML]{656565}{\color[HTML]{FFFFFF} \begin{tabular}[c]{@{}l@{}}Manufacturer \\ or designer\end{tabular}}} & \multicolumn{1}{l|}{\cellcolor[HTML]{656565}{\color[HTML]{FFFFFF} \textbf{Chip}}} & \multicolumn{1}{l|}{\cellcolor[HTML]{656565}{\color[HTML]{FFFFFF} Supports STS}} \\ \hline
{\color[HTML]{333333} Apple} & {\color[HTML]{333333} U1} & \checkmark\\ \hline
\rowcolor[HTML]{C0C0C0} 
{\color[HTML]{333333} Qorvo} & {\color[HTML]{333333} DW1000} &  \\ \hline
{\color[HTML]{333333} Qorvo} & {\color[HTML]{333333} DW3000 Family} & \checkmark\\ \hline
\rowcolor[HTML]{C0C0C0} 
{\color[HTML]{333333} NXP} & {\color[HTML]{333333} SR040/SR150} & \checkmark\\ \hline
{\color[HTML]{333333} NXP} & {\color[HTML]{333333} NCJ29D5} & \checkmark\\ \hline
\rowcolor[HTML]{C0C0C0} 
{\color[HTML]{333333} Imec} & {\color[HTML]{333333} ULP IR-UWB   radio} & \checkmark\\ \hline
\end{tabular}
\end{table}

The use of the STS field requires common knowledge of the keys and cryptographic scheme between the transmitter and receiver. Otherwise, decoding the STS fails, which causes the communication to fail. The way in which these keys are distributed between these devices is not specified in the standard. This problem is mostly agreed upon by higher layers and using a different wireless communication technology than UWB. UWB radio chips only supporting the base IEEE 802.15.4 standard are not capable of using the STS field. UWB radio chips supporting the IEEE 802.15.4z standard are required to support the STS, as it is essential for use cases where security is important, such as hands-free, location-aware keyless access. Due to the security requirements, such use cases are not supported by UWB radio chips only supporting the IEEE 802.15.4 standard. Using Table \ref{tab:STSsupport} there can be seen that the only chip that cannot be used for use cases that require the added security is the Qorvo DW1000 \cite{Firawhitepaper,Poturalski2011}.

\subsubsection{Modulation and encoding}
\begin{figure}[h!]
    \centering
    \includegraphics[width=0.48\textwidth]{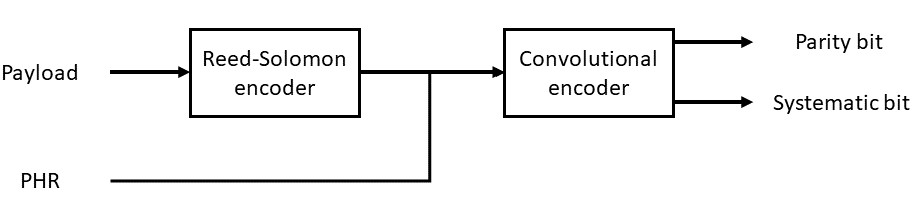}
    \caption{Payload and PHR encoding process (based on \cite{IEEE4}). }
    \label{fig:encodingscheme}
\end{figure}

In contrast to the preamble and SFD, the PHR and payload still need to be encoded and modulated. The encoding process is shown in Figure \ref{fig:encodingscheme}. The payload is first encoded using systematic Reed-Solomon block code. Figure \ref{fig:PHR} shows that the last 6 bits of the PHR field are used for SECDED encoding, therefore the Reed-Solomon encoding is omitted. Next, the PHR and payload are encoded using a convolutional encoder.
The IEEE 802.15.4 standard defines a half-rate convolutional encoder with K = 3. The IEEE 802.15.4z defines a new optional half-rate convolutional encoder with K=7. The systematic and parity bit generated by this process enable error detection and correction and are used in the modulation process.

\begin{figure}[h!]
    \centering
    \includegraphics[width=0.48\textwidth]{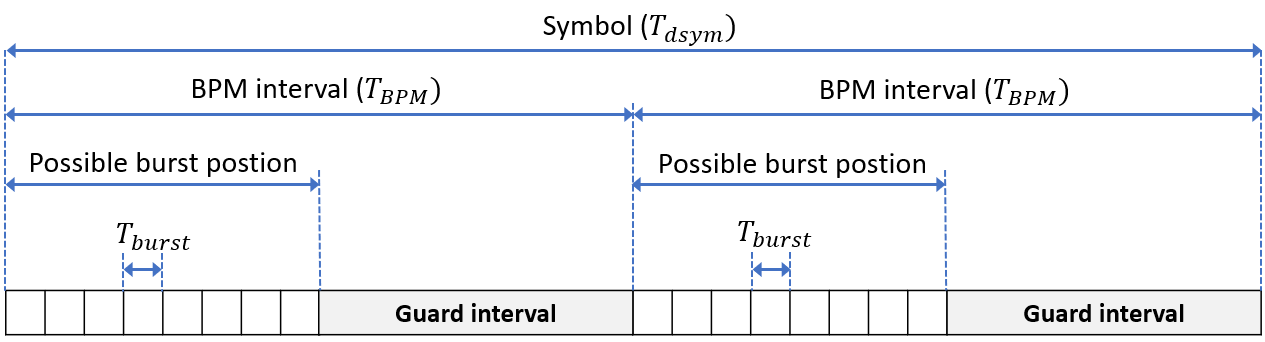}
    \caption{PHR and PHY payload symbol structure in the IEEE 802.15.4 standard (based on \cite{IEEE4}).}
    \label{fig:UWBsymbol}
\end{figure}

\begin{figure}[h!]
    \centering
    \includegraphics[width=0.48\textwidth]{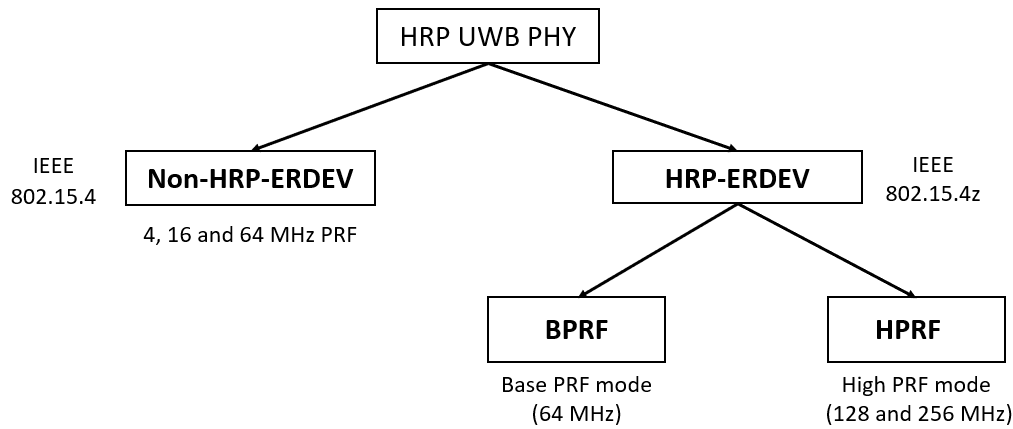}
    \caption{Overview of the different modes defined by the IEEE 802.15.4 and IEEE 802.15.4z standard (based on \cite{keysight})}
    \label{fig:PRFoverview}
\end{figure}

In UWB communication, a bit is transmitted using a train of pulses. The speed at which these pulses are sent is a parameter of the UWB system and is called the Pulse Repetition Frequency (PRF). The IEEE 802.15.4 HRP UWB  defines three options for the mean PRF: 4 MHz, 16 MHz, and 64 MHz. The IEEE 802.15.4 HRP UWB  defines a modulation scheme using BPM-BPSK depicted in Figure \ref{fig:UWBsymbol}. 
Each symbol is divided into two halves with duration $T_{BPM}= \frac{T_{dsym}}{2}$, this enables the BPM. Furthermore, each BPM interval is split in two halves: a possible burst position half and a guard interval which prevents interference from other systems that are sending. One burst is formed by $N_{cpb}$ pulses of length $T_c$. A burst can be sent in the first or second half of the symbol, this location indicates one bit and is determined by the systematic bit. The parity bit is transmitted using the phase of the burst: positive or negative.
The IEEE 802.15.4z standard drops support for PRFs of 4 and 16 MHz, and supports new higher PRFs of 128 and 256 MHz. The reason for these higher PRFs is again to increase the amount of threshold decision events and thus offer a higher dynamic range.
The 64 MHz PRF is combined with the same BPM-BPSK modulation as in the IEEE 802.15.4 standard and is called the Base Pulse Repetition Frequency (BPRF) mode. To enable the higher PRFs a new type of modulation is defined in IEEE 802.15.4z that only uses BPSK. The combination of a higher PRF and the BPSK modulation is called High Pulse Repetition Frequency (HPRF) mode. An overview of the different PRFs and modes supported by each standard is depicted in Figure \ref{fig:PRFoverview}. The IEEE 802.15.4z modes are called the HRP - Enhanced Ranging Device (HRP-ERDEV) modes and therefore the IEEE 802.15.4 standard is called the non-HRP-ERDEV.

\begin{table*}[]
\caption{Overview of supported modulation and PRF for each chip.}
\label{tab:PRFmod}
\centering
\begin{tabular}{|l|l|c|c|c|c|c|}
\hline
\rowcolor[HTML]{656565} 
\multicolumn{1}{|c|}{\cellcolor[HTML]{656565}{\color[HTML]{FFFFFF} \textbf{Manufacturer or designer}  }} & \multicolumn{1}{c|}{\cellcolor[HTML]{656565}{\color[HTML]{FFFFFF}  \textbf{Chip} }} & \multicolumn{3}{c|}{\cellcolor[HTML]{656565}{\color[HTML]{FFFFFF} BPM + BPSK}} & \multicolumn{2}{c|}{\cellcolor[HTML]{656565}{\color[HTML]{FFFFFF} BPSK}} \\
\cline{3-7} 
\rowcolor[HTML]{656565} 
\multicolumn{1}{|c|}{\multirow{2}{*}{\cellcolor[HTML]{656565}{\color[HTML]{FFFFFF}}}} & \multicolumn{1}{c|}{\multirow{2}{*}{\cellcolor[HTML]{656565}{\color[HTML]{FFFFFF} }}} & {\color[HTML]{FFFFFF} \begin{tabular}[c]{@{}c@{}}4MHz \\ PRF\end{tabular}} & {\color[HTML]{FFFFFF} \begin{tabular}[c]{@{}c@{}}16 MHz \\ PRF\end{tabular}} & {\color[HTML]{FFFFFF} \begin{tabular}[c]{@{}c@{}}64 MHz\\ PRF\end{tabular}} & {\color[HTML]{FFFFFF} \begin{tabular}[c]{@{}c@{}}128 MHz\\ PRF\end{tabular}} & {\color[HTML]{FFFFFF} \begin{tabular}[c]{@{}c@{}}256 MHz\\ PRF\end{tabular}} \\ \hline
{\color[HTML]{333333} Apple} & {\color[HTML]{333333} U1} &  ?&  ?& \checkmark& ? & ? \\ \hline
\rowcolor[HTML]{C0C0C0} 
{\color[HTML]{333333} Qorvo} & {\color[HTML]{333333} DW1000} & \checkmark& \checkmark& \checkmark&  &  \\ \hline
{\color[HTML]{333333} Qorvo} & {\color[HTML]{333333} DW3000 Family} &  & \checkmark& \checkmark&  &  \\ \hline
\rowcolor[HTML]{C0C0C0} 
{\color[HTML]{333333} NXP} & {\color[HTML]{333333} SR040/SR150} &  ?& ? & \checkmark& ?  & ?  \\ \hline
{\color[HTML]{333333} NXP} & {\color[HTML]{333333} NCJ29D5} &  ?& \checkmark& \checkmark& ? & ? \\ \hline
\rowcolor[HTML]{C0C0C0} 
{\color[HTML]{333333} Imec} & {\color[HTML]{333333} ULP IR-UWB} &  & \checkmark & \checkmark& \checkmark & \checkmark \\ \hline
\end{tabular}
\end{table*}

In Table \ref{tab:PRFmod} it can be seen that the HPRF mode is not mandatory, as the Qorvo DW3000 UWB radio chips supporting the IEEE 802.15.4z standard does not support the 128 and 256 MHz PRFs. For the Apple and NXP UWB radio chips, no information was found about the supported modulations. As the HPRF modes are not mandatory, it is not certain if they are supported. Because only IEEE 802.15.4z support is mentioned for these UWB radio chips, it is also not certain if 4 and 16 MHz PRFs are supported. For the NXP NCJ29D5 both IEEE 802.15.4 and IEEE 802.15.4z support is mentioned, this means that PRFs of 16 and 64 MHz are certainly supported by this chip.

In addition to the encoding scheme, modulation and PRF, the data rate needs to be identical as well to enable the receiver to decode the payload of the transmitted UWB frame and, thereby, enable communication and ranging. The data rates available for each chip depend on the supported PRFs.

\begin{table}[]
\centering
\caption{Supported data rates for each PRF in the HRP UWB in the IEEE 802.15.4/4z standards \cite{IEEE4,IEEE4z}.}
\label{tab:datarates}
\begin{tabular}{|c|c|c|c|c|}
\hline
\rowcolor[HTML]{656565} 
{\color[HTML]{FFFFFF} PRF} & {\color[HTML]{FFFFFF} 4 MHz} & {\color[HTML]{FFFFFF} \begin{tabular}[c]{@{}c@{}}16 MHz\\ 64 MHz\end{tabular}} & {\color[HTML]{FFFFFF} 128 MHz} & {\color[HTML]{FFFFFF} 256 MHz} \\ \hline
\cellcolor[HTML]{656565}{\color[HTML]{FFFFFF} \begin{tabular}[c]{@{}c@{}}Data rates\\ (Mbps)\end{tabular}} & \begin{tabular}[c]{@{}c@{}}0.11\\ 0.85\\ 1.70\\ 6.81\end{tabular} & \begin{tabular}[c]{@{}c@{}}0.11\\ 0.85\\ 6.81\\ 27.24\end{tabular} & 6.81 & 27.24 \\ \hline
\end{tabular}
\end{table}

In Table \ref{tab:datarates} it can be seen that the increased PRFs do not enable higher data rates. This is because the goal of the IEEE 802.15.4z standard was to enhance the IEEE 802.15.4 standard in terms of accuracy and integrity.

\subsection{LRP UWB PHY compatibility}

\subsubsection{LRP UWB PHY modes}
In the LRP UWB PHY, several modes are defined. A mode is defined by the combination of a modulation and PRF, which leads to a certain data rate. In Table \ref{tab:LRPmodes} an overview of all mode classes (a mode class is a category of modes with similar characteristics) is given together with the possible modulations, PRFs and data rates. The long-range, extended and base modes were defined in the IEEE 802.15.4 standard and the dual-frequency, extended dual-frequency and dual-frequency with enhanced payload capacity (EPC) were added in the IEEE 802.15.4z standard. The biggest change in the new modes is the use of dual-frequency. This is an extension to the OOK modulation where alternate OOK channels are used. The mode always transmits a pulse in either one of the two used frequency bands. The introduction of this dual-frequency causes the modulation to change from OOK to Pulsed-Binary-Frequency-Shift-Keying (PBFSK) or the combination of PBFSK with 8, 16, or 32 Pulse-Position-Modulation (PPM) for the EPC. A second change is an increased maximum PRF of 4 MHz.
\begin{table*}[]
\centering
\caption{Signaling modes and data rates for LRP UWB PHY in IEEE 802.15.4 and IEEE 802.15.4z \cite{IEEE4,IEEE4z}.}
\label{tab:LRPmodes}
\begin{tabular}{|c|c|c|c|c|c|}
\hline
\rowcolor[HTML]{656565} 
{\color[HTML]{FFFFFF} Mode class} & {\color[HTML]{FFFFFF} Modulation} & {\color[HTML]{FFFFFF} PRF} & {\color[HTML]{FFFFFF} Data rate} & {\color[HTML]{FFFFFF} IEEE 802.15.4} & {\color[HTML]{FFFFFF} IEEE 802.15.4z} \\ \hline
Long-range  & PPM & 2.0 MHz & 31.25 kbps & \checkmark& \checkmark\\ \hline
\rowcolor[HTML]{C0C0C0} 
Extended  & OOK & 1.0 MHz & 250 kbps & \checkmark& \checkmark\\ \hline
Base  & OOK & 1.0 MHz & 1 Mbps & \checkmark& \checkmark\\ \hline
\rowcolor[HTML]{C0C0C0} 
Dual-frequency modes & PBFSK & 1.0, 2.0, 4.0 MHz & 1, 2 Mbps &  & \checkmark\\ \hline
Extended dual-frequency modes & PBFSK & 1.0, 2.0, 4.0 MHz & \begin{tabular}[c]{@{}c@{}}250, 500 kbps\\ 1 Mbps\end{tabular} &  & \checkmark\\ \hline
\rowcolor[HTML]{C0C0C0} 
Dual-frequency with EPC modes & PBFSK-8/16/32PPM & 1.0, 2.0 MHz & 3, 4, 5, 6, 8, 10 Mbps &  & \checkmark\\ \hline
\end{tabular}
\end{table*}

\paragraph{Symbol structure}
Each mode class also has a different symbol structure. The symbol structure for each is explained in Table \ref{tab:LRPsymbol}. 

\begin{table}[h!]
\centering
\caption{Symbol structures of all LRP UWB PHY modes \cite{IEEE4,IEEE4z}.}
\label{tab:LRPsymbol}
\begin{tabular}{|l|l|}
\hline
\rowcolor[HTML]{656565} 
{\color[HTML]{FFFFFF}  Mode class} & {\color[HTML]{FFFFFF} Symbol structure} \\ \hline
Long-range  & \begin{tabular}[c]{@{}l@{}}Presence/absence of pulses in 1 MHz \\ PRF sequence\end{tabular} \\ \hline
\rowcolor[HTML]{C0C0C0} Extended  & \begin{tabular}[c]{@{}l@{}}Presence/absence of pulses in 1 MHz \\ PRF sequence generated by convolution \\ code with octal generators (5, 7, 7, 7).\end{tabular} \\ \hline
Base  & \begin{tabular}[c]{@{}l@{}}Manchester-encoded groups of 64 \\ pulses (32 on, 32 off) in 2 MHz PRF \\ sequence.\end{tabular} \\ \hline
\rowcolor[HTML]{C0C0C0} \begin{tabular}[c]{@{}l@{}}Dual-frequency \\ modes\end{tabular} & \begin{tabular}[c]{@{}l@{}}Presence of pulses at either one of the \\ center frequencies  transmitted at \\ nominal PRF values of 1, 2 or 4 MHz\end{tabular} \\ \hline
\begin{tabular}[c]{@{}l@{}} Extended \\ dual-frequency \\ modes \end{tabular} & \begin{tabular}[c]{@{}l@{}}Four pulses per symbol generated by \\ convolution with octal generators \\ (5, 7, 7, 7)\end{tabular} \\ \hline
\rowcolor[HTML]{C0C0C0}\begin{tabular}[c]{@{}l@{}} Dual-frequency \\ with EPC modes\end{tabular} & \begin{tabular}[c]{@{}l@{}}Presence of pulses at either one of the \\ center frequencies transmitted at \\nominal PRF values of 1 or 2 MHz with \\ modulation of the pulse position within \\ the symbol using 8, 16 or 32 PPM\end{tabular} \\ \hline
\end{tabular}
\end{table}

\paragraph{Frame structure}
The frame of the LRP UWB PHY start with a preamble. The used preamble depends on the mode class. The different preambles are shown for all mode classes in Table \ref{tab:LRPpreamble}.

\begin{figure}[h!]
    \centering
    \includegraphics[width=0.49\textwidth]{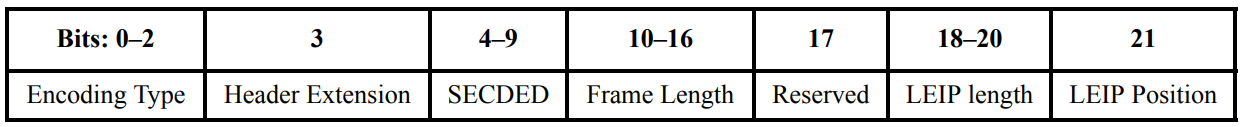}
    \caption{PHR format for LRP UWB PHY in IEEE 802.15.4 \cite{IEEE4}).}
    \label{fig: LRPPHR}
\end{figure}

\begin{table}[]
\centering
\caption{Preamble generation of all LRP UWB PHY modes \cite{IEEE4,IEEE4z}.}
\label{tab:LRPpreamble}
\begin{tabular}{|l|l|}
\hline
\rowcolor[HTML]{656565} 
{\color[HTML]{FFFFFF} Mode class} & {\color[HTML]{FFFFFF} Preamble} \\ \hline
Long-range & \begin{tabular}[c]{@{}l@{}}3 segments: (1) continuous stream\\ of pulses at 2 MHz PRF, (2) a \\ pulse/no-pulse sequence, (3) \\ series of between 16 and 64 “1” \\ symbols.\end{tabular} \\ \hline
\rowcolor[HTML]{C0C0C0} 
Extended & \begin{tabular}[c]{@{}l@{}}Continuous stream of pulses at\\ 1 MHz PRF, length between \\ 16-256.\end{tabular} \\ \hline
Base & \begin{tabular}[c]{@{}l@{}}Continuous stream of pulses at \\ 1 MHz PRF, length between\\  16-256.\end{tabular} \\ \hline
\rowcolor[HTML]{C0C0C0} 
Dual-frequency & \cellcolor[HTML]{C0C0C0} \\ \cline{1-1}
Extended dual-frequency & \cellcolor[HTML]{C0C0C0} \\ \cline{1-1}
\rowcolor[HTML]{C0C0C0} 
Dual-frequency with EPC & \multirow{-3}{*}{\cellcolor[HTML]{C0C0C0}\begin{tabular}[c]{@{}l@{}}Continuous stream of pulses with\\  alternate binary values at the PRF\\  specified in the mode.\end{tabular}} \\ \hline
\end{tabular}
\end{table}
After the preamble, the SFD is transmitted. In the IEEE 802.15.4 standard, there was only one LRP UWB SFD code with a length of 16. The IEEE 802.15.4z standard defines additional SFD codes with a length of 32, 64 or 128.
The next part of the frame is the PHR, this field is shown in Figure \ref{fig: LRPPHR}. 

The PHR consists of the following fields:
\begin{itemize}
    \item \textbf{Encoding type}: indicates the symbol mapping and encoding that is used.
    \item \textbf{Header extension}: if this bit is set, the payload is discarded.
    \item \textbf{SECDED}: Hamming block code to enable single error correction and two error detection at the receiver.
    \item \textbf{Frame length}: integer set to the length of the payload.
    \item \textbf{Reserved}: Reserved for future use (indicates ranging in IEEE 802.15.4z)
    \item \textbf{LEIP length}: indicates the length of the Location Enhancing Information Postamble (LEIP).
    \item \textbf{LEIP position}: specifies the position of the optional LEIP sequence.
\end{itemize}

The only difference between the two standards in the PHR is that the reserved field is used for ranging in the IEEE 802.15.4z \cite{4zLRPppt}.

The payload follows the PHR. For all modes except the dual-frequency with EPC mode, encoding of the payload is the same as the other fields and thus as explained in Table \ref{tab:LRPsymbol}. The EPC mode provides higher data rates in the payload and inserts a guard interval to accommodate high RF multipath. The symbol consists of an active interval where PPM is used and an inactive guard interval.

The LEIP is an optional postamble. This field consists of a sequence of UWB pulses at the PRF of the mode  that is used to enhance the ability to locate the
transmitter.
\subsubsection{Ranging}
Ranging support is added in the IEEE 802.15.4z standard. This is done by adding a basic ranging scheme using Round-Trip Time-of-Flight (RTToF). This is done using fixed Receive-to-Transmit turnaround time. Devices that are capable of this (supporting the IEEE 802.15.4z standard) know when a fixed turnaround time is necessary using the “reserved” bit in the PHR, other devices will just ignore it \cite{4zLRPppt}.

\subsubsection{Conclusion}
The previous sections show that the LRP UWB PHY in IEEE 802.15.4 and IEEE 802.15.4z are only compatible for communication when the long-range, extended or base mode is used. This means that the improved data rates, sensitivity and power consumption from the IEEE 802.15.4z standard are not available. They are also not compatible for ranging following the standard. However, the company 3dB access had already implemented ranging similarly using LRP UWB before the IEEE 802.15.4z standard was released. 

\section{MAC layer compatibility}
\label{sec:MAClayer}
\subsection{MAC layer}
The MAC layer is one of the two sub-layers that make up the Data Link layer of the OSI model. This layer defines protocols to allow for different UWB systems to use the same channel. The IEEE 802.15.4 standard defines a MAC layer, and the IEEE 802.15.4z provides enhanced MAC functionality based on this standard. An introduction to the MAC frame format defined in the IEEE 802.15.4/4z standard is given below \cite{IEEE4,IEEE4z}.

\subsubsection{General MAC message format}
\begin{figure}[h!]
    \centering
    \includegraphics[width=0.48\textwidth]{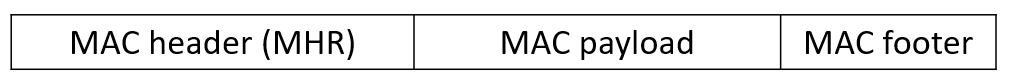}
    \caption{General MAC frame (based on \cite{IEEE4}).}
    \label{fig: MACframe}
\end{figure}

The MAC message (or frame) fills the payload portion of the UWB PHY frame, as depicted in Figures \ref{fig:structure-4} and \ref{fig:structure-4z}. The MAC frame is composed of a header, followed by a payload of variable length and finally ends with the MAC footer, as shown in Figure \ref{fig: MACframe}. 

\begin{figure}[h!]
    \centering
    \includegraphics[width=0.48\textwidth]{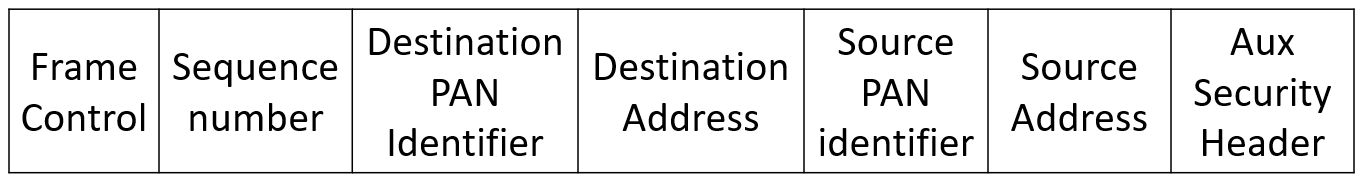}
    \caption{The MAC header (based on \cite{IEEE4}).}
    \label{fig: MACheader}
\end{figure}

The MAC footer is a Frame Checking Sequence (FCS) cyclic redundancy check (CRC) that is used to detect transmission errors. Figure \ref{fig: MACheader} shows the MAC header, used to identify a frame, in more detail. For example, the destination address is used to filter the frames that are destined for the receiver.

The frame control field is a 16-bit field that starts all IEEE 802.15.4/4z frames. The purpose of this field is to indicate the frame type and which components are part of the MAC header.

This frame consists of the following subfields:
\begin{itemize}
    \item \textbf{The Frame type field} specifies the type of frame using 3 bits. The possible frame types are Beacon, Data, Acknowledgement, MAC command, Multipurpose and Fragment. 
    \item \textbf{The Security enabled field} indicates if the Auxiliary Security Header field is used in the MAC header, using 1 bit.
    \item \textbf{The Frame pending field} specifies if the sender has more data for the receiver.
    \item \textbf{The Acknowledgement request field} uses 1 bit to indicate if the receiver needs to acknowledge the received frame.
    \item \textbf{The Personal Area Network (PAN) ID compression field} uses 1 bit to indicate whether the MAC frame contains only one of the PAN identifier fields, even though both source and destination addresses are present in the MAC frame. 
    \item \textbf{The Destination addressing mode field} indicates the presence and size of the destination address using 2 bits.
    \item \textbf{The Frame version field} is used to specify the version number of the frame. This is necessary because the frame was changed in the 2003 version of the IEEE 802.15.4 standard. 
    \item \textbf{The Source addressing mode field} is used to indicate the presence and size of the source address using 2 bits.
\end{itemize}

As the MAC frame has not changed in IEEE 802.15.4z enhancement of the MAC, there are no consequences for compatibility when UWB radio chips use the different standards. The biggest enhancement to the MAC is the addition of some localization techniques in the functional description. Before, it was completely up to the manufacturer/designer to define the localization technique. More information about UWB localization techniques is provided in \ref{subsec:rangingschemes}.

\subsection{Multiple access schemes}
Due to the different physical properties of UWB compared to narrowband wireless technologies, different multiple access schemes need to be used \cite{Gupta2007}. While the different UWB radio chips can use the same MAC frame format, the lack of consensus on which multiple access schemes are best for UWB systems causes all UWB systems to use proprietary  multiple access schemes as no standard multiple access scheme is defined. Chip suppliers, like Qorvo and NXP, leave the implementation of the MAC layer to the host microprocessor system controlling the chip. Companies selling complete UWB systems and consumer products using UWB implement a proprietary MAC layer that is not released to the public.This means that compatibility of the multiple access scheme is only possible if developers of UWB systems share which multiple access scheme they use. 

\section{Localization techniques}
\label{subsec:rangingschemes}
UWB technology allows for accurate timing on the arrival of the signal, however the main use case of the UWB technology is localization. For this, the distance or the relative position between two UWB devices is needed. This is calculated from the timing on the signal using a localization technique. The most used localization techniques are discussed below, using a UWB tag and multiple UWB anchors. The goal is to determine the location of the tag.

\subsection{ToF} 
\begin{figure}[ht]
    \centering
    \includegraphics[width=0.45\textwidth]{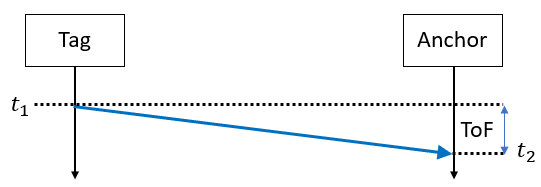}
    \caption{ToF localization technique.}
    \label{fig:ToF}
\end{figure}

This method uses the propagation time to calculate the distance between the tag and anchor nodes, as depicted in Figure \ref{fig:ToF}. The tag transmits a UWB frame with as payload, the time at which the frame is sent ($t_1)$. The anchor receives the frame at $t_2$ and calculates the ToF as $t_2-t_1$. The signals are electromagnetic and travel at the speed of light ($c = 299.8 * 10^6 m/s.$), therefore the range is found using $d = c*ToF$. When the distance between three anchor nodes and the tag has been calculated, the location of the tag can be determined using trilateration. The drawback of this method is that precise synchronization between all nodes is necessary. The precision of this synchronization has a direct impact on the accuracy of the ranging \cite{Shi2016}.

\subsection{TDoA}
\begin{figure}[ht]
    \centering
    \includegraphics[width=0.45\textwidth]{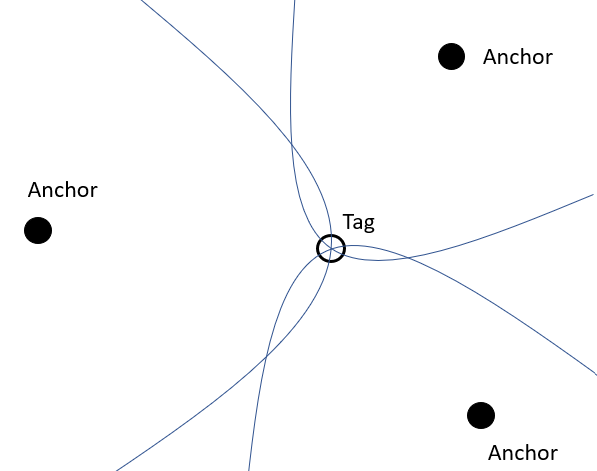}
    \caption{TDOA localization using hyperbolas.}
    \label{fig:TDOA}
\end{figure}

The tag will send out a signal, which will arrive at all anchors at a different time, due to the anchors being at different distances from the tag. The difference between the arrival of the signal in two anchors can be used to calculate a hyperbola. The intersection of at least three hyperbolas gives the location of the tag, as depicted in Figure \ref{fig:TDOA}. The tag itself will never know its position, unless it is transmitted back. Whether the tag needs to know its position depends on the application. For example, in an automotive hands-free access control application the car needs to know the distance with the key, but the key does not need to know that information. Important to note is that while the tag does not need to be synchronized with the anchors, the anchors must be synchronized with each other \cite{Shi2016}.
\subsection{TWR}
\begin{figure}[ht]
    \centering
    \includegraphics[width=0.49\textwidth]{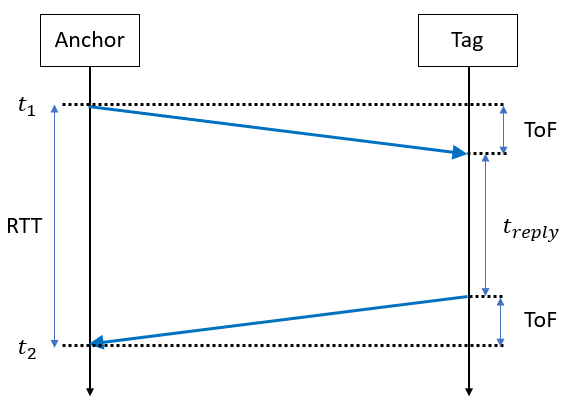}
    \caption{The message transmissions in TWR.}
    \label{fig:TWR}
\end{figure}

This method is an improvement on the ToF method, which eliminates the need for synchronization between the anchor and tag. This is achieved by only using timestamps from one device. The anchor transmits a message that is received at the tag after the propagation time or ToF. The tag responds after a fixed reply time. This reply time is included in the packet to calculate ToF from the Round-Trip Time (RTT) at the anchor. This is depicted in Figure \ref{fig:TWR}. This RTT can be used to calculate the distance between tag and anchor. When three anchors perform this TWR, the location of the tag can be determined using trilateration. A variant of this is Double-Sided TWR (DS-TWR) where at least three messages are transmitted instead of only two for TWR. This approach has the advantage that both anchor and tag can calculate the distance between them \cite{TWR}.

\subsection{Consequences for compatibility}
The use of localization techniques is important for compatibility, as both sender and receiver need to transmit the correct and necessary frames to calculate the distance and or location. The use of the localization technique mostly depends on the use case of the technology. TDoA is used in applications where the tag does not need to know its own location, like  asset tracking and other RTLS products. TWR is used in ad-hoc, non-permanent applications of UWB like hands-free access control.
For compatibility to be possible, the UWB sender and receiver need to agree upon the localization technique that is used. As chip suppliers like Qorvo and NXP leave the implementation of the MAC layer to the host microprocessor system controlling the chip, these chips can be configured to use all possible localization techniques. This is done because the choice of localization technique is highly dependent on the configuration, the system design, and application requirements. In the FiRa and Apple standard, there is negotiation between transmitter and receiver on their capabilities. The transmitter chooses, and the receiver can be informed by side channel or higher level information. The problem for compatibility can be that commercial systems implement proprietary localization techniques or proprietary ways to decide upon which technique to use. The IEEE 802.15.4z standard adds the description of some localization techniques (TWR, DS-TWR, TDOA and ToF) to the MAC functional description which indicates that these are the recommended techniques that should be available on devices supporting the IEEE 802.15.4/4z standards.

\section{Device Discovery compatibility}
\label{sec:DeviceDiscovery}
Before communication between two UWB devices can start, device discovery needs to be performed. 
Device discovery is a process where UWB devices carry out a search to find other UWB devices to communicate with. There are several standards that define how this device discovery can be implemented. Due to the energy consuming nature of UWB radio chips compared to other wireless technologies, most of these standards rely on a secondary channel (often a Bluetooth radio) to discover nearby UWB devices.

\subsection{FiRa standard}
\begin{figure}[h!]
    \centering
    \includegraphics[width=0.48\textwidth]{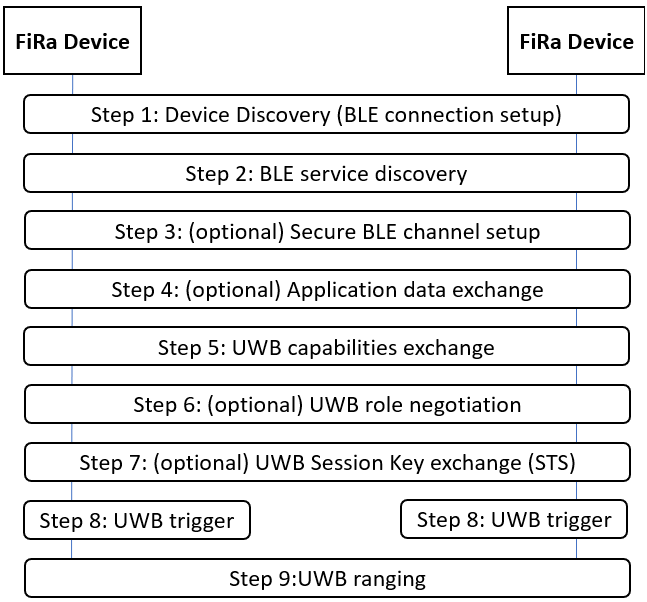}
    \caption{Device discovery and ranging setup procedure in the FiRa standard (based on \cite{FiRaCSML}).}
    \label{fig:FiRadiscovery}
\end{figure}

In Figure \ref{fig:FiRadiscovery}, the device discovery and ranging setup procedure from the FiRa Common Service Management Layer (CSML) is depicted. The first step in the procedure is the device discovery using an out-of-band channel, typically Bluetooth Low Energy (BLE) but potentially NFC or other wireless technologies. Once two UWB devices have discovered each other using BLE, the BLE service discovery is performed and optionally a secure BLE channel is set up, and application data is exchanged. Then the UWB capabilities are exchanged and the UWB parameters are decided upon. After optionally negotiating the UWB role and session key exchange, the UWB system is triggered and the UWB ranging is started \cite{FiRaCSML}.

\begin{figure}[h!]
    \centering
    \includegraphics[width=0.46\textwidth]{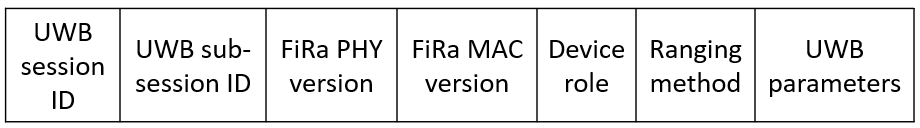}
    \caption{the FiRa UWB configuration message (based on \cite{FiRaCSML}).}
    \label{fig:firaconfiguration}
\end{figure}

\begin{itemize}
    \item During the procedure, the UWB capabilities are exchanged over the out-of-band channel (BLE, NFC, …) using a RESTful interface in the form of the UWB\_CAPABILITY message. Figure \ref{fig:firacapability} shows this message.
    
This RESTful message contains the following information: FiRa PHY version, FiRa MAC version, Device Roles and lastly UWB parameter support. This last field consists of the following subfields: multi-node support, STS configuration support, Ranging methods support, Ranging Round Hopping, Supported channels, RFRAME feature capability, extended MAC address, short MAC address, UWB initiation time, AoA support, Block Striding Capability, Ranging Time Structure support, Scheduled Mode support, Device Class, PRF Mode support, Convolutional code length support, List of BPRF parameters set supported, List of HPRF parameters set supported \cite{FiRaCSML}.

\begin{figure}[ht]
    \centering
    \includegraphics[width=0.46\textwidth]{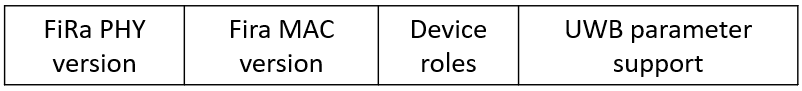}
    \caption{The FiRa UWB capability message (based on \cite{FiRaCSML}).}
    \label{fig:firacapability}
\end{figure}

    \item After the exchange of UWB\_CAPABILITY, the chosen UWB\_CONFIGURATION is decided upon. To this end, the configuration messages are similarly exchanged over the same out-of-band channel (BLE, NFC, …) using a RESTful interface. The UWB\_CONFIGURATION message is shown in \ref{fig:firaconfiguration}. This message contains following information: UWB session ID, FiRa PHY version, FiRa MAC version, Ranging method and UWB parameters. The UWB parameters field contains the following subfields: Multi-node mode, RFRAME configuration, STS configuration, Round Hopping, Scheduled mode, Contention phase length, Ranging time structure, Block striding, Ranging interval, Responder slot index, Channel number, Preamble code index, PRF mode, Ranging frequency, slot duration, …
\end{itemize}

\subsection{Apple Nearby Interaction}
\begin{figure}[h!]
    \centering
    \includegraphics[width=0.48\textwidth]{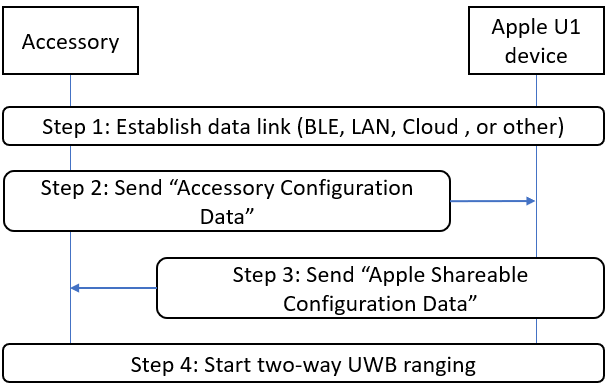}
    \caption{Device discovery and ranging setup between an accessory and an Apple device (based on \cite{AppleNIprotocol}).}
    \label{fig:AppleNIprocedure}
\end{figure}

Figure \ref{fig:AppleNIprocedure} depicts the device discovery and ranging setup between an accessory and an Apple device containing the U1 UWB chip. First, discovery is performed using a different technology than UWB. In contrast to the FiRa standard, this discovery is not limited to BLE. Discovery and setup of a data link can be performed using different methods like, LAN, Cloud, …

\begin{figure}[ht]
    \centering
    \includegraphics[width=0.49\textwidth]{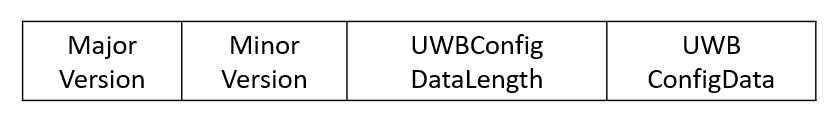}
    \caption{The Apple Shareable Configuration Data message \cite{AppleNIprotocol}.}
    \label{fig:accessoryconfigdata}
\end{figure}

Step two consists of the accessory generating and sending the 'Accessory Configuration Data'. This message format is shown in Figure \ref{fig:accessoryconfigdata} and consists of several parameters:
\begin{itemize}
    \item \textbf{Major Version}: must match between devices. The only defined major version at this moment is 1.
    \item \textbf{Minor Version}: must match between devices. Only defined minor version at this moment is 0.
    \item \textbf{Preferred Update Rate}: Accessory must select a preferred update rate. The options are automatic, infrequent and user interactive. When automatic is selected the Apple device will select the update rate, when infrequent is selected the update rate will be approximately once per second and when user interactive is selected the update rate is on the scale of 5 per second.
    \item \textbf{Reserved for future use}.
    \item \textbf{UWB Configuration Data Length}: specifies the length of the Configuration Data field.
    \item \textbf{UWB Configuration Data}: shall be provided by the UWB middleware to the
embedded application through a dedicated interface that is not further specified in the Nearby Interaction Accessory Protocol Specification \cite{AppleNIprotocol}.
\end{itemize}

\begin{figure}[ht]
    \centering
    \includegraphics[width=0.49\textwidth]{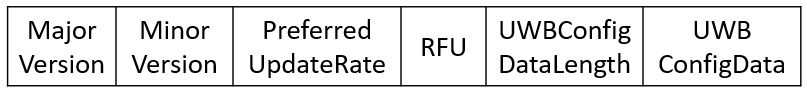}
    \caption{The Apple Accessory Configuration Data message \cite{AppleNIprotocol}.}
    \label{fig:shareableconfigdata}
\end{figure}

Step three consists of the Apple U1 device generating and sending the 'Apple Shareable Configuration Data'. This message format, shown in Figure \ref{fig:shareableconfigdata}, is similar to the 'Apple Accessory Configuration Data' message, but some fields are omitted.
The last step is to set up the UWB ranging using the parameters from the UWB Configuration Data fields \cite{AppleNIprotocol}.
\subsection{Car Connectivity Consortium (CCC)}
No details of the CCC digital key release 3.0 have been published at the time of writing. 
\subsection{Consequences for compatibility}
The device discovery approach in the different standards is similar. First discovery is performed using a different wireless communication technology than UWB, most commonly BLE. Next, the different UWB nodes negotiate the UWB settings that will be used. Finally, the UWB connection is set up using the UWB settings that the nodes agreed upon. 
Despite these similarities, the different standards are not interoperable as the message that are transmitted during the discovery procedure are not the same. Apart from these standards, device discovery can also be implemented in a proprietary way by different companies that provide UWB systems.

\section{Future research trends and directions}
\label{ResearchDirections}
\subsection{Antenna design challenges}
While we have mainly focused on the interoperability between UWB radio chips from different vendors, it has been demonstrated in literature that inappropriate design of the transmit and receive antenna may lead to severe orientation-specific pulse distortion and undesired phase-center variations, thereby adversely affecting IR-UWB RTLS performance \cite{Mahfouz2008,Sibille2005,Boryssenko2006,Qing2006,Tan2013} and also potentially endangering interoperability. As a consequence, conventional frequency-domain-based figures-of-merit, such as return loss and gain radiation pattern, do no longer suffice to characterize IR-UWB antennas. To accurately predict system-level performance/compatibility, a new set of metrics is required. The system fidelity factor (SFF) was introduced in \cite{Quintero2011} to characterize the amount of pulse distortion introduced by the antenna system. Furthermore, in \cite{Vandenbrande2021}, the Distance Estimation Error (DEE) was proposed to characterize the amount of ranging bias. Moreover, with UWB localization systems entering the stage of mass production and mass integration in a wide variety of heterogeneous IoT environments, where IR-UWB antennas are invisibly and compactly integrated within the object or onto the person that needs to be positioned, special care should be devoted to considering the antenna integration platform. Hence, stand-alone antenna design in free-space conditions does no longer suffice. UWB antenna system design should rather focus on guaranteeing the desired performance in the envisaged deployment scenario by considering the influence of the integration platform. However, no commercial simulation tools currently exist to efficiently and simultaneously optimize for frequency-domain and system-level antenna metrics over the antenna’s field of view. Moreover, different UWB antenna vendors use different system-level antenna metrics to quantify the (orientation-specific) pulse distortion. This makes accurate and complete IR-UWB RTLS design very challenging.
Therefore, future research should focus on a holistic system-level optimization framework that jointly optimizes conventional antenna-oriented parameters and relevant system-level figures-of-merit, while considering integration platform effects. In parallel, the IEEE standard for Definitions of Terms for Antennas \cite{Roederer2014} should be extended with these relevant system-level figures of merit to facilitate comparison between IR-UWB antennas from different vendors. Finally, with the advent of IR-UWB-based AoA estimation techniques, leveraging multi-antenna systems for the accurate and precise extraction of AoA information, a similar exercise is needed to (1) identify and define a relevant set of system-level figures of merit for such UWB multi-antenna systems (such as the differential group delay versus the AoA, as proposed in \cite{Dotlic2018}) besides the more conventional antenna-array-oriented figures-of-merit (embedded element pattern, active s-parameters, …) and (2) to efficiently optimize for these system-level figures of merit.

\subsection{PHY layer challenges}
\subsubsection{Improving on the IEEE 802.15.4z standard}
The need of a follow-up on IEEE 802.15.4z is motivated by the fact that the application of UWB has expanded rapidly and has become part of high-volume consumer platforms. It is being applied to an ever-wider range of applications using the unique capabilities of UWB to provide very accurate ranging, localization, sensing and data communication with excellent coexistence properties. New applications require flexibility and scalability in network typology's, varying in size, shape and number of devices from a few devices within a meter or less of each other to hundreds or more devices up to 100 m distant. Expanding data rates available to both lower rates with greater distances than current rates, and higher rates at short distances. This expands the options for trading distance, range and energy consumption. 

For these purposes, IEEE Task Group 15.4ab “Next Generation UWB Amendment” \cite{nextgen802} has been created. The objectives are enhancements to 802.15.4 UWB PHY and MAC and associated ranging techniques while retaining backward compatibility with ERDEVs.

Possible enhancements include: additional coding, preamble and modulation schemes to additional coding, preamble and modulation schemes to support improved link budget and/or reduced air-time relative to IEEE 802.15.4z UWB; additional channels and operating frequencies; interference mitigation techniques to support greater device density and higher traffic use cases relative to the IEEE 802.15.4z UWB; improvements to accuracy, precision and reliability and interoperability for high-integrity ranging; schemes to reduce complexity and power consumption; definitions for tightly coupled hybrid operation with narrowband signaling to assist UWB; enhanced native discovery and connection setup mechanisms; sensing capabilities to support presence detection and environment mapping; and mechanisms supporting low-power low-latency streaming as well as high data-rate streaming allowing at least 50 Mb/s of throughput. Support for peer-to-peer, peer-to-multi-peer, and station-to-infrastructure protocols are in scope, as are infrastructure synchronization mechanisms. This amendment includes safeguards so that the high throughput data use cases do not cause significant disruption to low duty-cycle ranging use cases.

The cut-off date for new PHY proposals was May 2022, and for new MAC proposals July 2022. The targeted standard date is end 2023 / beginning 2024.  

\subsubsection{Standardization of UWB AoA}
AoA is an interesting technique for UWB localization, as the location of a tag can be estimated using a single anchor, equipped with at least two antennas, by combining AoA with a distance measuring method. In the localization techniques mentioned in Section \ref{subsec:rangingschemes} the location of the tag can only be determined using multiple anchors. There are several AoA methods available \cite{Dotlic2018}, a few examples are mentioned here:
\begin{itemize}
    \item ToF method: the difference in ToF measurement for the two antennas at the receiver can be used to calculate the AoA.
    \item TDoA: the difference in arrival time for the same frame is used to estimate the AoA
    \item Phase Difference of Arrival (PDoA): the difference in phase of the received carrier is used to estimate the AoA.
\end{itemize}
Several recent UWB radio chips can calculate the AoA, like Qorvo DW1000, Qorvo DW3000, NXP SR150, imec ULP IR-UWB radio and Apple U1. Unfortunately, no standard has incorporated AoA estimation in the specification. This means that for AoA implementation, each UWB system can implement its own proprietary AoA method. This is due to the AoA method heavily depending on the implementation of the antenna, which in turn is influenced by the equipment design. Future research could focus on several aspects such as (i) defining a standardized AoA estimation method in the PHY layer standards, (ii) negotiating about the possibilities for AoA in the UWB device discovery standards and (iii) defining common data representations for exchanging angle information. 

\subsubsection{UWB radar standardization}
Before, we mostly focused on the localization use case of UWB technology. However, the technology has different applications as well, one of them being radar. For example, UWB radar can be used for human presence and activity detection \cite{human-presence1,human-presence2,li2012through}, … and health monitoring: non-contact heart rate and respiratory rate determination \cite{radarvitalsigns,immoreev2008uwb,staderini2002uwb}, … While the UWB radar use case seems promising, it has not yet been added to any UWB standard. However, they still need to fulfill at least the spectrum mask requirements defined for UWB technology.
Future research could be performed to define a standardized UWB PHY for UWB radar. This standardization could enable commercial use of UWB radar technology.

\subsubsection{Pulse shape}
While both the IEEE 802.15.4 and IEEE 802.15.4z standard have the same requirements for the pulse shape, it was found that differences in pulse shape between different UWB radio chips are still possible. Further research into the influence of the pulse shape on the performance of UWB systems could be performed. In this research, the influence of a different pulse width between two different UWB radio chips can be investigated. The result can be used to find a way to mitigate possible ranging errors caused by the difference in pulse width, this could allow for more accurate ranging between two different UWB radio chips.

\subsection{Data link layer challenges}

\subsubsection{Standardization of the MAC protocol}
The goal of the MAC layer is to trigger, schedule and share measurement results, for the efficient gathering of information in a scalable and low power manner. As mentioned in the MAC layer overview, the MAC frame formats are standardized, but the way these frames are exchanged are not. In scientific literature, multiple MAC protocols for UWB have been proposed, ranging from uncoordinated MAC protocols for localization (ALOHA based) to synchronized time division multiple access (TDMA) based MAC protocols \cite{UWBMACloc,MACmultiple,BAUWENS2021102637}. As a result, no commercial localization systems are currently interoperable, thus requiring different user tags for each building that is entered. There is thus a strong need for a standard that can discover the type of MAC protocol (synchronized, non-synchronized) that is supported by previously deployed infrastructure nodes  as well as the supported configuration (user roles, duration of the superframe, network join process, etc.).

In Section \ref{sec:DeviceDiscovery} it was explained that both the FiRa and Apple standard use a different wireless communication technology than UWB for device discovery. In \cite{Gupta2007} and \cite{Karapistoli2012} it is shown that traditional MAC protocols are not suitable for UWB networks. Combining this information could imply that narrowband systems are intrinsically more suited for some MAC functions than UWB. Future research could be performed to determine if this is true and in which situations this is the case and why. It might be that hybrid systems, as they appear in the FiRa and Apple standard, are more desirable in some situations.

\subsubsection{Performance analysis of device discovery approaches}

Although the FiRa and Apple standards support device discovery, a thorough analysis and comparison in terms of overhead, latency and scalability of these two standards is still lacking. The analysis can show the influence of choices, made during the design of the standards, on these different standards. This analysis can help in making the decision of which standard will be adopted in new UWB systems.

\subsubsection{Link configuration decision algorithm}

While FiRa and Apple define messages to exchange the supported PHY layer configurations, they do not define any decision algorithms that define which settings should be selected. While UWB performs very well in open spaces and line-of-sight (LOS) conditions, accuracy can rapidly degrade in NLOS and crowded environments. However, good accuracy is possible in more difficult environments when using specific configurations. A possibility for mitigating this problem is developing a decision algorithm that determines the best configurations for a UWB link using the available UWB capabilities of both UWB radio chips and the available link estimation parameters. To enable this in a way that ensures compatibility, a few subcomponents need to be in place:
\begin{itemize}
    \item A standardized UWB capabilities exchange format.
    \item A standardized format for exchanging link state measurements used to determine the best configurations in that link state.
    \item Decision algorithm that determines the best configuration, considering the available UWB capabilities, the link state measurements and the application requirements (expected accuracy, expected ranging distance, maximum latency, maximum energy consumption, etc.). This algorithm will use the available information to handle channel allocation, power control and interference management.
    \item A standardized protocol to enable the configuration determined by the decision protocol. 
\end{itemize}

Developing a well-functioning decision algorithm is particularly important, as selecting the wrong configurations can have a major negative impact. In the future, different techniques for implementing the decision algorithm can be researched and compared. Even though the decision algorithm is the key component, without developing a compatible format and protocol for the capabilities exchange and adaptation of configurations, this cannot be adopted in the most common UWB systems.

\subsection{Application layer challenges}

\subsubsection{Standardized data formats}

Up until now, we focused on the possibility to communicate / range between UWB radio chips. Assuming that the process works, application developers will need a standardized way to interpret the system output information such as distances, positions, and angles between UWB devices. 

There is already a wide range of global organizations that provide standardized information representations and semantics for global interoperability in IoT networks. Some examples include oneM2M \cite{onem2m}, the Open Mobile Alliance (OMA) \cite{OMA} and the Internet Protocol for Smart Objects (IPSO) Alliance (merged into OMA). For example, the authors of \cite{s18072142} define Lightweight M2M (LwM2M) Position Object Models for representing position information. However, most information models of these standardization bodies currently focus on sensor data rather than positioning information. No standardized representation is currently defined for advanced UWB output such as angles, distances, etc., making it challenging for application developers to design cross-platform and cross-system user applications.

In terms of standardization of position information, recently, an alliance named Omlox \cite{omlox} has defined an open standard for RTLS systems. With this standard, various localization technologies, for example UWB, Wi-Fi, BLE, GPS, RFID and 5G, can be easily connected. This standard tries to enable localization compatibility between different localization systems by introducing a common way to exchange distances for the different technologies. Similarly, relative UWB position information can be converted to global GPS coordinates. A similar application layer standard for only UWB ranging could be developed to enable RTLS without the need for compatibility on lower levels. 

Similar standardized information models will permit multiple localization systems
to communicate and interoperate with each other in order to obtain better context information and resolve positioning errors or conflicts.

\subsubsection{RTLS standards}
Currently, upper layer standards, like the FiRa standard and Apple Nearby Interaction protocol, are focused on the device-to-device application of UWB technology of which the best known is access control. However, they do not define how the standard can be extended to RTLS applications. RTLS technology allows for location tracking of individuals or objects with high accuracy within buildings, such as warehouses, campuses, and hospitals to improve inventory management. To this end, the FiRa and Apple negotiation protocols do not only need to negotiate about PHY layer configurations, but would also need to define which localization approaches are supported. Future research to help define a standard RTLS protocol could be performed. By performing this research, Apple and/or the FiRa consortium could be stimulated to add support for RTLS and in this way enable compatibility between different UWB radio chips for RTLS purposes.

\subsubsection{Smartphone compatibility}
Besides Apple, other prominent smartphone makers have released devices that contain UWB radio chips as well. Samsung has released the Galaxy Note 20 Ultra and Galaxy S21 containing an UWB chip from NXP \cite{nxpsamsung} and Xiaomi announced the release of the MIX4 smartphone also containing an NXP UWB chip \cite{xiaominxp}. Google has added an UWB API to Android 12, however this API is part of the System APIs. This means that the API is currently unavailable to third-party apps. At this moment it is not clear if the API is part of the System APIs because it is not yet ready for full release or because Google wants to limit the use of the UWB technology deliberately. An analysis could be made of the difference between the UWB API available for the Android operating system compared to the UWB API (Nearby Interaction Protocol) available on iOS. This analysis can then find out what the influence is of the possible different choices that have been made in the design of the two APIs, and if future compatibility and interoperability could be possible.

\subsection{UWB regulations}
The use case limitations for UWB (outdoor, aviation, …) are not globally harmonized and therefore, UWB regulation for outdoor usage can differ from country to country. For example, permanently installed outdoor UWB systems are prohibited in most countries and regions. As the UWB technology matures, permanent outdoor UWB systems are becoming more attractive. To allow for these applications of UWB technology to be developed, research could be performed to help define globally harmonized regulations for outdoor UWB usage.

\section{Conclusion}
\label{sec:conclusion}
This paper provides a comprehensive overview of the different standards that are defined for UWB communication. While previous papers \cite{Sedlacek2019,Firawhitepaper, singh2021security} focused on the enhancements in security and accuracy from the IEEE 802.15.4 to IEEE 802.15.4z standard, this paper focuses on the implications for compatibility at the PHY layer and the MAC and upper layers. For each of these layers, an overview of the different standards that are defined for that specific layer is given, as well as the consequences for the compatibility that the differences between the standards have.

\begin{itemize}
    \item \textit{PHY compatibility}: PHY compatibility between UWB radio chips supporting the IEEE 802.15.4 and IEEE 802.15.4z standard is possible if the correct settings are configured. First, the same channels need to be selected on both UWB radio chips. Currently, channel 5 is supported on all UWB radio chips from the market overview in Section \ref{sec:market}. Next, the same preamble and SFD codes need to be selected. Only the ternary preamble codes of length 127 and the ternary SFD code of length 8 are supported by both standards. This means that the higher accuracy enabled by the new codes in the IEEE 802.15.4z standard is not available in compatibility mode. The same frame and PHR structure need to be configured, this means that the higher security provided by the STS field as well as the increased payload length provided by new optional PHR structures is not available in compatibility mode. The BPRF mode, using the BPM-BPSK modulation and 64 MHz PRF, is the only compatible mode between the two standards. This means that the higher accuracy and better balance between airtime per data bit and the number of pulses per data bit provided by the new modulations and higher PRFs from the IEEE 802.15.4z standard is not available in compatibility mode. Finally, the pulse shape requirements are the same in both standards, but there was found that this does not mean that the pulse shape of different UWB radio chips is identical. A difference in pulse width can influence the ranging accuracy as the timing on the pulses, and thus ranging distance, can differ. 
    \item \textit{MAC layer and localization technique compatibility}: Although MAC frame structures are standardized, the implementation of the multiple access scheme as well as the localization technique is mostly proprietary. Chip suppliers, like Qorvo and NXP, leave the implementation of the MAC layer and localization technique to the host microprocessor system controlling the chip. The use of the same MAC and localization technique is important for the compatibility and interoperability, as both sender and receiver need to transmit the necessary frames at the correct times to calculate the distance and or location.
    \item \textit{Device Discovery compatibility}: Chip suppliers, like Qorvo and NXP, can leave the implementation of device discovery to the host microprocessor system controlling the chip. However, there are some standards defined that handle this procedure that are also supported by the UWB radio chips from these suppliers. FiRa has defined the FiRa CSML, Apple the Nearby Interaction Protocol and the CCC the Digital Key Release 3.0. The procedure in the different standards is similar. First discovery is performed using a different wireless communication technology than UWB, most commonly BLE. Next, the UWB settings are negotiated, and finally the connection is set up. Despite the similarities of the process from FiRa and Apple, compatibility between the standards is not possible as the transmitted message during the procedures are different. Moreover, no algorithms are defined to determine the optimal settings based on channel conditions, application requirements and supported configurations.
    
 As such, it was shown that either the IEEE 802.15.4 or the IEEE 802.15.4z standard for the PHY layer is supported by each chip. As mentioned, there is compatibility between these two standards. However, the situation at the higher layers is more complicated, as UWB systems can implement proprietary approaches or use one of the standards defined for these layers. They need to have the same MAC, ranging and device discovery procedures configured for the communication to be available. 
 
 The data sheets of the different UWB devices are a first indication that interoperability can be possible. However, it is not enough to guarantee interoperability. For this, actual communication test results or certification is necessary. The NXP SR040/SR150 have base FiRa certification \cite{firacertified}. Qorvo is a sponsor and board member of FiRa and the DW3000 family of UWB radio chips are developed in accordance with the FiRa consortium PHY and MAC specification, but there is no official FiRa certification at the time of writing. For the Apple U1 chip, interoperability with the NXP SR040/SR150 and Qorvo DW3000 is available \cite{DW3000apple,NXPapple} and there exists development kits available for development of UWB enable applications between the Apple U1-devices and the NXP SR040/SR150 or Qorvo DW3000 UWB radio chips.
 
 Finally, the paper identified and described a number of future research directions and standardization challenges, such as extending current PHY standards to also support angle of arrival data, defining link estimation and decision algorithms for PHY layer setting configurations, defining common data formats and representations to exchange position and distance information, standardizing the MAC protocols and extending the current device discovery standards to also support RTLS systems. 
\end{itemize}
\bibliographystyle{IEEEtran}
\bibliography{sample-base}
\begin{IEEEbiography}[{\includegraphics[width=1in,height=1.25in,clip,keepaspectratio]{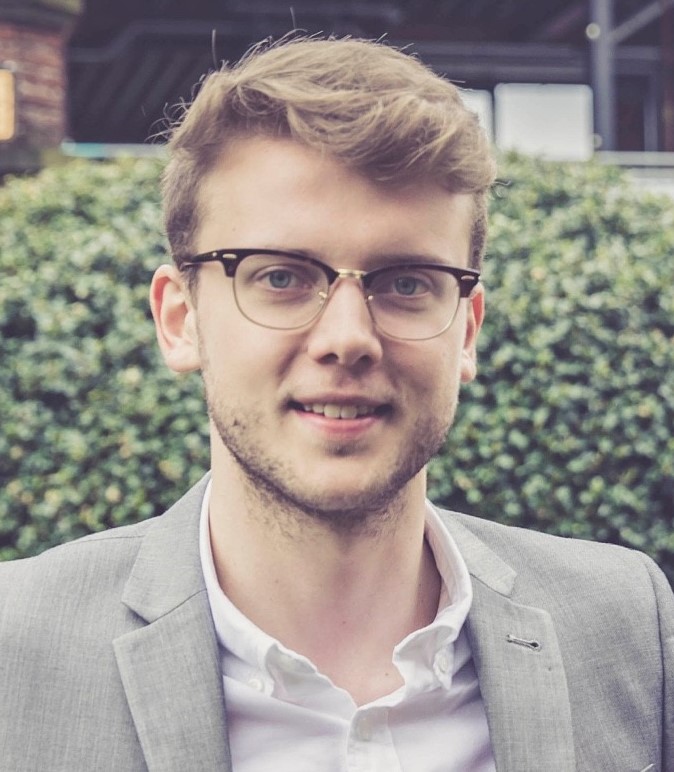}}]{Dieter Coppens}
received the master's degree in electrical engineering from Ghent University, Belgium, in 2021. He is currently pursuing the Ph.D. degree with the IDLab Research Group. His research interests are wireless networking, indoor localization system based on Ultra-wideband technology and machine learning.
\end{IEEEbiography}
\begin{IEEEbiography}[{\includegraphics[width=1in,height=1.25in,clip,keepaspectratio]{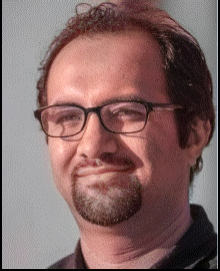}}]{Dr. Adnan Shahid} (Senior Member, IEEE) received the B.Sc. and M.Sc. degrees in Computer Engineering from the University of Engineering and Technology, Taxila, Pakistan, in 2006 and 2010, respectively, and the Ph.D. degree in Information and Communication Engineering from Sejong University, South Korea in 2015. From 2015 till 2016, he worked as an Assistant Professor at Taif University, Kingdom of Saudi Arabia. He is currently working as Senior Researcher at IDLab, which is a research group of imec within Ghent University and University of Antwerp. He is currently technically leading the Erasmus+ DigiHealth-Asia project (2021-2024) and has been the technical manager of the European Space Agency CODYSUN project. He has been involved in several projects: DARPA Spectrum Collaboration Challenge (SC2), European H2020 research projects (eWINE and WiSHFUL), and national projects (SAMURAI, IDEAL-IOT, and Cognitive Wireless Networking Management). He is Secretary and voting member of IEEE P1900.8 - Standard for Training, Testing, and Evaluating Machine-Learned Spectrum Awareness Models. He is author/coauthor of more than 70 publications in well-known journals and conferences. His research interests include resource optimization, interference management, self-organizing networks, small cell networks, machine learning, artificial intelligence, IoT, localization and 5G/xG networks. He is also an Associate Editor in journals, like IEEE Access and Journal of Networks and Computer Application (JNCA), Elsevier.
\end{IEEEbiography}
\vfill
\begin{IEEEbiography}[{\includegraphics[width=1in,height=1.25in,clip,keepaspectratio]{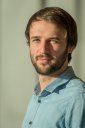}}]{Prof. dr. ing. Sam Lemey} (Member, IEEE) received the M.Sc. degree in electronic engineering from Howest, University College West Flanders, Kortrijk, Belgium, in 2012 and the Ph.D. degree in electrical engineering from Ghent University, Ghent, Belgium, in 2016. He is currently professor at the Department of Information Technology (INTEC), Ghent University/imec. From January to March 2018, he was a visiting scientist with the Terahertz Photonics Group, Institute of Electronics, Microelectronics and Nanotechnology (IEMN), University Lille Nord de France, Lille, France. His research interests include antenna systems for wearable applications, active antenna design for the Internet of Things and (beyond) 5G applications, (opto-electronic) millimeter-wave multi-antenna systems, impulse-radio ultra-wideband antenna systems for centimeter-precision localization, and full-wave/circuit co-optimization frameworks to realize (opto-electronic) active (multi-)antenna systems. Sam Lemey received the URSI Young Scientist Award at the 2020 URSI General Assembly and was awarded the Best Paper Award at the 2016 IEEE MTT-S Topical Conference on Wireless Sensors and Sensor Networks. He was also a co-recipient of the 2015 Best Paper Award at the 22nd IEEE Symposium on Communications and Vehicular Technology in the Benelux and of the 2019 ECOC Best Demo Award.
\end{IEEEbiography}
\begin{IEEEbiography}[{\includegraphics[width=1in,height=1.25in,clip,keepaspectratio]{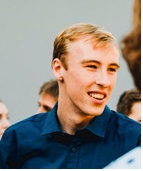}}]{Ben Van Herbruggen}
 was born in Antwerp in 1995. He received the M.Sc. degree in electrical engineering from Ghent University, Belgium, in July 2018. In September 2018, he started as Research Assistant with the Department of Information Technology (INTEC) at Ghent University in the IDLab research group towards a PhD degree. His scientific work is focused on the use of energy harvesters for wireless networking and indoor localization system based on Ultra-wideband technology. Ben is the author and co-author of various publications on Ultra-wideband localization solutions.
\end{IEEEbiography}
\begin{IEEEbiography}[{\includegraphics[width=1in,height=1.25in,clip,keepaspectratio]{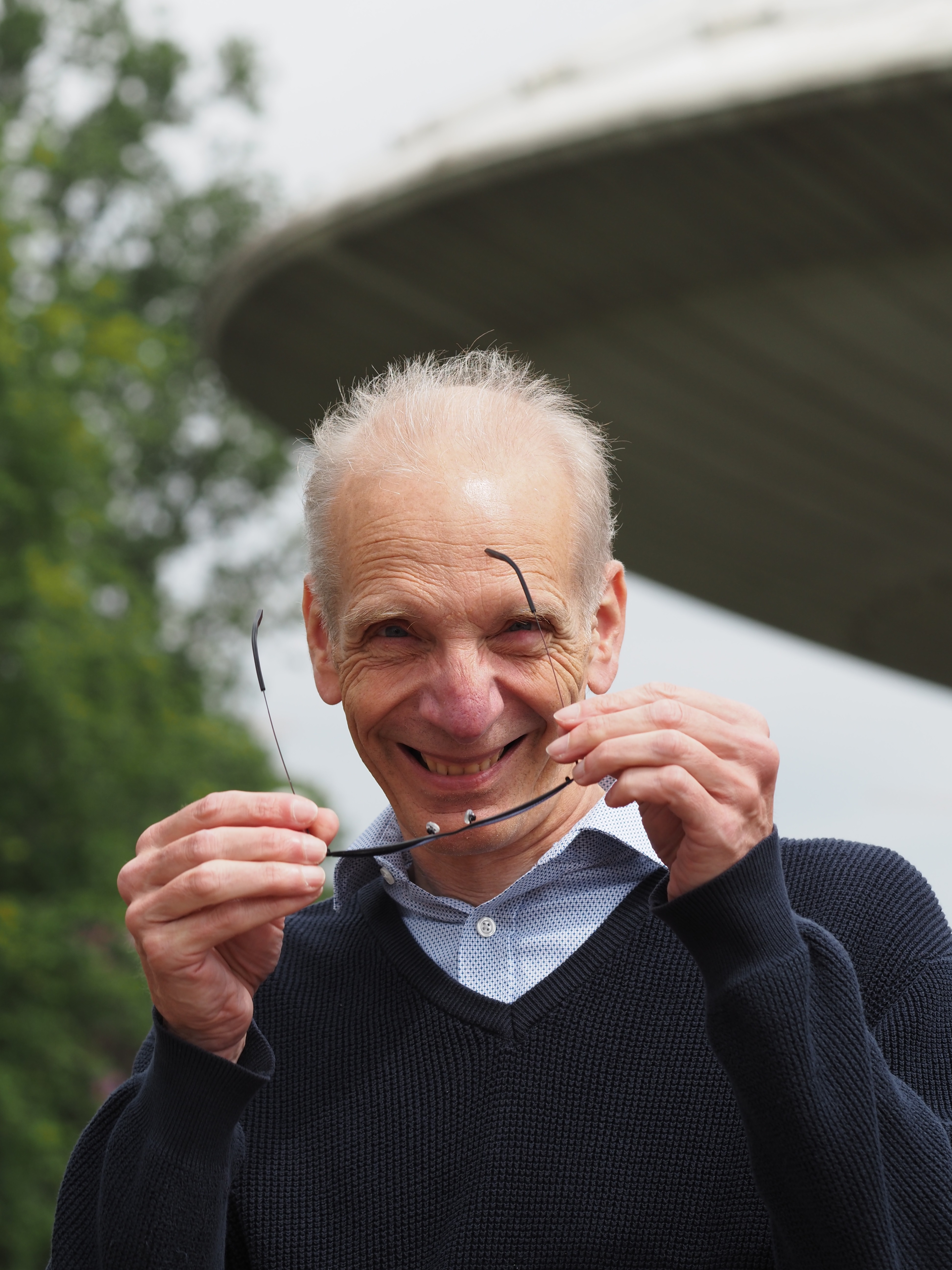}}]{Dr. Chris Marshall} graduated from Cambridge University, UK in 1980 and holds an MA in Natural Sciences and received a PhD in electrical engineering from Imperial College London.  He has worked as a scientist, engineer and group leader with Philips Research in the UK and with Philips Kommunikations Industrie AG and Philips Semiconductors Technology Centre for Mobile Communications in Nuremberg, Germany, at the leading edge of the integration of consumer wireless communications devices, with pioneering wireless designs for paging, DECT and GSM, and the creation of the low power radio system that became standardized in 802.15.4.  He led the development of the first software radio GPS receiver, and was Chief Technical Officer of the spin-out Geotate, which created software and services for geotagging photographs, and was acquired in 2009 by u-blox.  At u-blox Dr. Marshall developed new systems for positioning and synchronization using cellular signals.  He is named inventor for some 50 patents, an Honorary Senior Research Fellow at Sussex University, UK, and is now working with imec in the Netherlands, developing Bluetooth and UWB solutions for IoT positioning.
\end{IEEEbiography}
\begin{IEEEbiography}[{\includegraphics[width=1in,height=1.25in,clip,keepaspectratio]{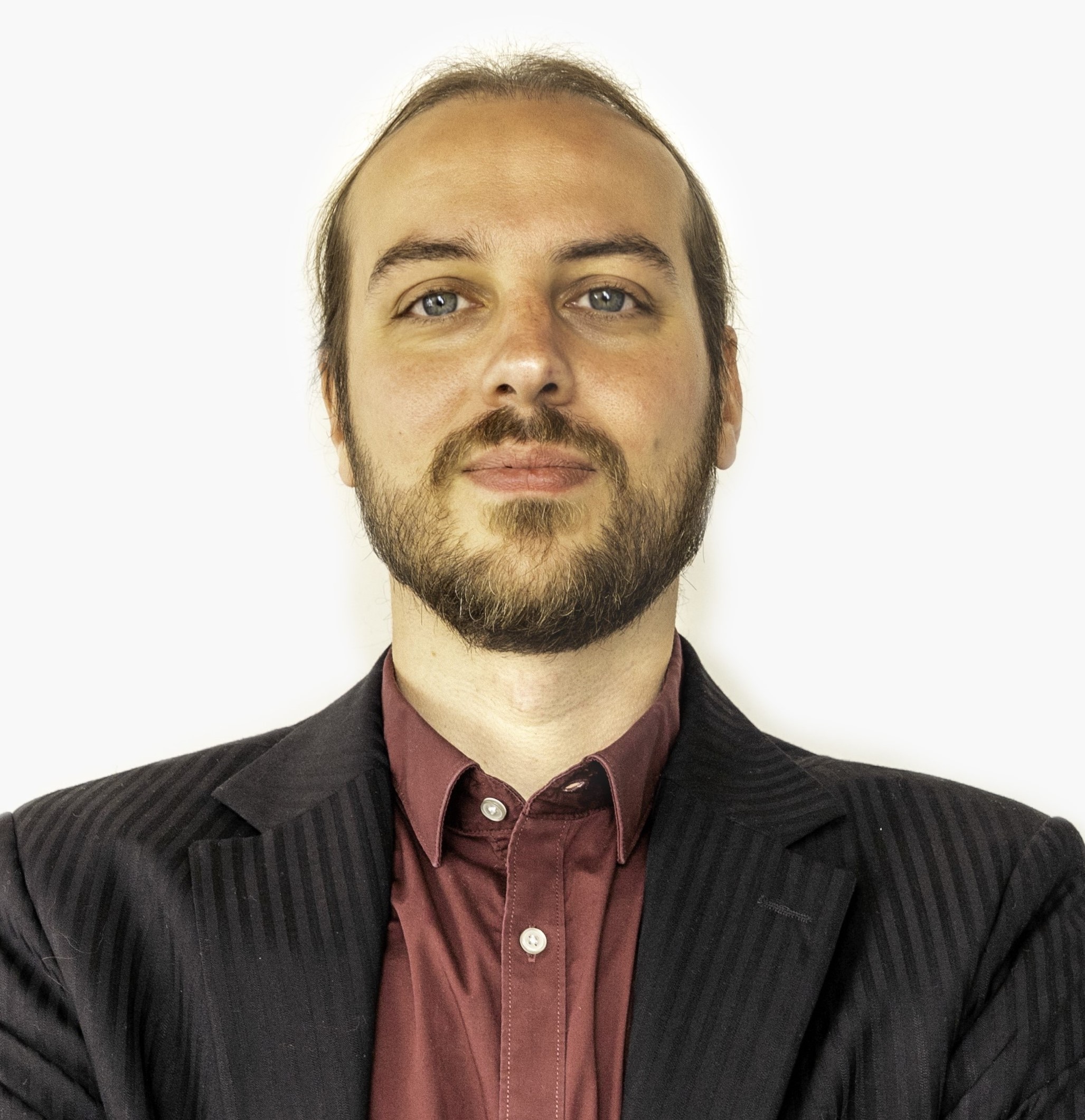}}]{Prof. dr. ir. Eli De Poorter} is professor at the IDLab research group from imec and Ghent University. His team performs research on wireless communication technologies such as (indoor) localization solutions, wireless IoT solutions and machine learning for wireless systems. He performs both fundamental and applied research. For his fundamental research he is currently the coordinator of several research projects (SBO, FWO, GOA, etc.) and has over 200 publications in international journals or in the proceedings of international conferences. For his applied research, he collaborates with industry partners to transfer research results to industrial applications, as well as to solve challenging industrial research problems. He is also co-founder of the lopos spin-off company (https://lopos.be) which offers privacy-aware UWB wearables for safety and social distancing.
\end{IEEEbiography}
\vfill
\EOD
\end{document}